%% file: main.tex
\documentclass[a4paper,fleqn,usenatbib]{mnras}

\usepackage[T1]{fontenc}
\usepackage{ae,aecompl}

\usepackage{graphicx}	% Including figure files
\usepackage{amsmath}	% Advanced maths commands
\usepackage{amssymb}	% Extra maths symbols

\title{Type II supernovae Early Light Curves}

\author[Shussman, Waldman \& Nakar]{Tomer Shussman$^{1}$\thanks{Contact e-mail: \href{shussman@post.tau.ac.il}{shussman@post.tau.ac.il}}, Roni Waldman$^{2,3}$ and Ehud Nakar$^{1}$
\\
$^{1}$The Raymond and Beverly Sackler School of Physics and Astronomy, Tel Aviv University, Tel Aviv 69978, Israel \\
$^{2}$Racah Institute of Physics, The Hebrew University, Jerusalem 91904, Israel \\
$^{3}$Particle Physics \& Astrophysics Dept., Weizmann Institute of Science, Rehovot 76100, Israel}

\date{Last updated \today}

\pubyear{2016}

\begin{document}
\label{firstpage}
\pagerange{\pageref{firstpage}--\pageref{lastpage}}
\maketitle

\begin{abstract}
Observations of type II supernova early light, from breakout until recombination, can be used to constrain the explosion energy and progenitor properties. Currently available for this purpose are purely analytic models, which are accurate only to within an order of magnitude, and detailed numerical simulations, which are more accurate but are applied to any event separately. In this paper we derive an analytic model that is calibrated by numerical simulations. This model is much more accurate than previous analytic models, yet it is as simple to use. To derive the model we analyze simulated light curves from numerical explosion of $124$ red supergiant progenitors, calculated using the stellar evolution code MESA. We find that although the structure of the progenitors we consider varies, the resulting light curves can be described rather well based only on the explosion energy, ejecta mass and progenitor radius. Our calibrated analytic model, which is based on these three parameters, reproduces the bolometric luminosity within $25\%-35\%$ accuracy and the observed temperature within $15\%$ accuracy (compared to previous analytic models which are indeed found to be accurate only to within an order of magnitude). We also consider deviations of the early time spectrum from blackbody, and find that the Rayleigh-Jeans regime is slightly shallower (roughly $L_\nu \propto \nu^{1.4}$). This modified spectrum affects the optical/near-UV light curve mostly during the first day when the typical observed temperature is $\gg 10^4 ~^\circ$K. We use our results to study the optical and near-UV early light curves from first light until recombination and briefly discuss what can be learned from current and future observations. Light curves generated using our calibrated model can be downloaded at \url{http://www.astro.tau.ac.il/~tomersh/}.
\end{abstract}

%%%%%%%%%%%%%%%%%%%%%%%%%%%%%%%%%%%%%%%%%%%%%%%%%%

%%%%%%%%%%%%%%%%% BODY OF PAPER %%%%%%%%%%%%%%%%%%

\section{Introduction}
\label{section:introduction}
Type II-P and possibly also type II-L supernova (SN) are generated by the explosion of red supergiants (RSGs) \citep{Smartt2009}. In these progenitors, the core collapse generates a shock which propagates through the hydrogen-rich envelope. As the shock reaches the stellar surface, first light is emitted \citep{Colgate1974, Falk1978, Imshennik1981, Ensman1992, Matzner1999}. After the shock breaks out of the star, the envelope radiates as it expands, leading to a long lasting emission that decays slowly \citep{Grassberg1971, Chevalier1976, Chevalier1992, Tominaga2009, Piro2010, Nakar2010, Rabinak2011, Dessart2013}.

Observations of type II SN light curves can be used to constrain the stellar properties and explosion energy, and shed light on the inner structure of RSG progenitors. Observations at early times (until about 10 days after the breakout) may be especially useful since the physics at these times is relatively "clean" from recombination, line emission, and radioactive decay, as pure hydrodynamic and radiation transport in ionized hydrogen govern the evolution and emission. In addition, early observations can provide independent constraints on the explosion and progenitor parameters. Finally, the emission at early times allows for probing the outer parts of the progenitor ($10^{-3}-10^{-1}M_\odot$) which are otherwise hardly accessible. During the last decade, growing numbers of type II SNe early light curves became available \citep[e.g.,][]{Gezari2008, Schawinski2008, Arcavi2012, Anderson2014, Faran2014, Sanders2015, Gonzales2015, Gall2015, Rubin2015}. In the near future, we expect that multi-wavelength observations of the early emission will be widely available. Inspired by the possibility of future observations, we revisit the subject.

Previous studies of the light curve generated by a shock breakout and cooling envelope emission at early times were either based on numerical calculations of specific progenitors \citep[e.g.,][]{Shigeyama1988, Woosley1988, Ensman1992, Blinnikov1998, Schawinski2008, Tominaga2009, Tominaga2011, Dessart2013, Morozova2016} or on using analytic models \citep[e.g.,][]{Weaver1976, Chevalier1992, Matzner1999, Piro2010, Katz2010, Sapir2011, Sapir2016}, in which a specific ("analytic") progenitor profile was assumed. Recently, \citet{Nakar2010} [NS10] and \citet{Rabinak2011} [RW11] proposed analytic models to describe the bolometric luminosity and observed temperature of RSG progenitors from first light to about $10$ days.

The main advantage of the analytic models is that they provide global relations between the observables and the explosion and progenitor properties, which can be easily applied to large data sets, such as the ones that have recently begun accumulating. In addition, the models enable predictions of the signal based on the progenitor properties, for the planning of future observations. Finally, the use of analytic relations enhances the understanding of the underlying physical mechanisms. However, these models are limited by the need to make many approximations on the stellar structure, the dynamics of the problem and the radiation transfer. The result of these approximations is that the analytic models are accurate only to within an order of magnitude. Numerical models, on the other hand, require less assumptions and are therefore more accurate, but so far they were used to study specific SNe or to generate light curves for a set of progenitors, without providing a general model for the effect of each parameter on the early light curve.

The goal of this paper is to provide an analytic model, with all of its advantages, that is accurate at a level close to that of numerical simulations. For that we numerically simulate the explosions from a large set of progenitors, and combine the results with analytic understanding of the light curve evolution, in order to construct a calibrated analytic model that is simple, yet accurate, and can be used to analyze large data sets in the future.

In order to do so, we separate the problem into two stages. First, we calculate numerically the light curves generated by exploding progenitors with the same density profile, but with different explosion energies, and progenitor radii and masses. We consider an analytic progenitor prototype with a structure similar near the stellar edge to that assumed in previous analytic studies (e.g., NS10, RW11), but more realistic in inner parts of the star. For these progenitors, the effect of each progenitor and explosion property on the light-curve is extracted for a wide range of values, independently of other properties. Since the early emission depends mostly on the conditions at the breakout in the outer parts of the progenitor, and not directly on the total explosion energy and ejecta mass, we find the dependence of the light curve on properties of the breakout (e.g., breakout velocity, density at the breakout location, etc.). We also find a mapping between the breakout parameters and the global ones (explosion energy, ejecta mass and progenitor radius) to obtain the dependence of the light curve on these parameters. Second, we use a large set of more than a $100$ RSG progenitors, calculated using the stellar evolution code MESA, to study the effect of different, more realistic structures on the emission. Here we also find the dependence of the light curve on the breakout parameters and then use them to find the dependence on the global SN parameters. Despite the large set of numerically calculated progenitors, their features are limited to the range of progenitor parameter space that we study and the specific schemes used in the stellar evolution simulations. Therefore the relations obtained based on this set are more accurate but are specific in part for the progenitors we checked. We discuss which of our results are more general and which depends more strongly on the exact progenitor structures. We also pay special attention to the deviation of the observed spectrum from a blackbody in the Rayleigh-Jeans regime and construct an analytic approximation of the observed spectrum.

The paper proceeds as follows: A brief review of the theory developed in previous analytic studies is presented in section \ref{section:theory}. In section \ref{section:semi_analytic} we obtain relations between progenitor and explosion properties and the bolometric luminosity and observed temperature for an analytic star. We show where earlier analytic models are accurate, where they fail and why. Results for numerically calculated stars are shown and discussed in section \ref{section:numerical}, where the effects of deviation of the observed spectrum from a blackbody and of light travel time are also studied. In section \ref{section:final} we summarize the model which relates the light-curve to the breakout and progenitor properties. A reader that is interested only in the final scaling relations should refer to this section. An analysis of the optical light curve appears in section \ref{section:optical}, and an analysis of the velocity of the photosphere appears in section \ref{section:photosphere}. Our main conclusions are summarized in section \ref{section:discussion}.

\section{Theory}
\label{section:theory}

SN explosion drives a radiation dominated shock that propagates through the decreasing density profile of the stellar envelope. At first (after the shock crosses envelope mass that is comparable to the He core mass), the dynamics of the problem are similar to the Sedov-Taylor \citep{Taylor1950, Sedov1959} explosion, but as the shock reaches the edge of the star, it accelerates because of the steep density decrease. A hydrodynamic solution for shock acceleration was proposed by \citet{sakurai1960}, for a density profile $\rho \propto (R_* - r)^n$ where $R_*$ is the progenitor radius, $r$ is the distance from the center and $n$ is a parameter which is usually assumed to equal $n=1.5$ for a RSG. This solution is planar and is therefore applicable only to the stellar edge, namely, $R_*-r \ll R_*$. Being mediated by radiation, the shock has a width of optical depth $\tau_{\rm{s}} \simeq c/v_{\rm{s}}(r) \simeq c/v(r)$ \citep{Weaver1976}, where $c$ is the speed of light and $v_{\rm{s}}(r)$ [$v(r)$] is the shock [matter just behind the shock] velocity. The shock keeps accelerating up to the point at which the optical depth for photons to escape the envelope $\tau(r)$ becomes comparable to $\tau_{\rm{s}}$. At this point, the shock "breaks-out" of the star, and first radiation is emitted. After the breakout, the envelope keeps radiating as it expands. During the expansion, a rarefaction wave propagates inwards and the velocity of the outer layers of the envelope roughly doubles itself. A model that describes the hydrodynamic expansion of the envelope is presented by \citet{Matzner1999}.

The hydrodynamic evolution of the expansion has two phases - a planer phase, which occurs at early times, when the expanding gas radius has not yet doubled its initial radius, and a spherical phase which begins approximately when the radius is doubled. The nature of the emitted radiation changes significantly between the phases. During the planar phase $\tau$ of each mass element is constant while the diffusion time from each element grows linearly in time, similarly to the dynamical time. Therefore during this phase radiation escapes only from the same mass shell from which radiation escaped at the breakout (NS10, \citealt{Piro2010}). Following NS10, we denote this shell the breakout shell, and note that it obeys $\tau \simeq c/v_{\rm{s}}$ at the breakout. During the spherical phase, however, $\tau$ of each mass element decreases with time, and inner shells begin to dominate the emitted radiation. Through the entire evolution (from breakout and up to recombination) the observed luminosity is generated at the shell that satisfies $\tau \approx c/v$. This shell is denoted (following NS10) as the luminosity shell.

During the expansion the envelope cools down rapidly because of adiabatic and radiation losses. At around $t \approx 10 - 20$ days after the breakout, the observed temperature reaches $T \approx 7500^{\circ}$K, and recombination of hydrogen atoms becomes significant. The recombination yields a rapid opacity drop which affects the observed temperature and luminosity. In addition, at times earlier than $t \approx 10 - 20$ days, the contribution of deposited energy by $^{56}\rm{Ni}$ radioactive decay to the light curve is negligible.

In this work, we focus on calculating the emitted radiation at early times, when the temperature is high enough for the hydrogen to be fully ionized and $^{56}\rm{Ni}$ radioactive decay is negligible. For these times, NS10 obtained an analytic estimate of the emitted radiation at the breakout, and the planar and spherical phases. In order to do that, they used the following approximations and assumptions:

\begin{description}
\item[Assum. 1] The explosion is fully spherical, and therefore the evolution is one dimensional.
\item[Approx. 2] Energy deposition by $^{56}\rm{Ni}$ radioactive decay is negligible.
\item[Approx. 3] Radiation transport is treated in the diffusion approximation. This is justified since both the luminosity and the observed temperature are determined at optical depth $\tau > 1$.
\item[Approx. 4] When enough photons can be generated to maintain thermal equilibrium, the spectrum is assumed to be a blackbody. 
\item[Approx. 5] The diffusion opacity is dominated by Thomson scattering (e.g. over absorption processes) of fully ionized hydrogen and helium with primordial ratios, and the absorption opacity, which is used to determine the observed temperature, is dominated by  free-free absorption (e.g. over bound-free and bound-bound transitions).
\item[Approx. 6] The progenitor density profile near the edge is of a power law form with $n=1.5$. (which is typical for RSG progenitors).
\item[Assum. 7] The emission is always determined by shells near to the edge, namely, with initial (pre-explosion) coordinates $R_* - r \ll R_*$.
\item[Assum. 8] The breakout shell structure during the expansion is of a simple planar rarefaction wave\footnote{Prior to the breakout, the evolution is governed by hydrodynamics alone, everywhere except for the breakout shell, where radiation transfer is important. Therefore, it is harder to describe it analytically.}.
\end{description}

NS10 found that the bolometric luminosity consists of two power-laws, corresponding to the two phases of evolution. Defining $t$ as the time relative to the bolometric peak emission, the luminosity scales as $t^{-\alpha}$ where $\alpha_{\rm p}=4/3$ for the planar stage, and $\alpha_{\rm{s}}=0.17$ for the spherical stage (equation 29 in NS10). The planar power law origins in the fact that only the breakout shell radiates during the planar stage, while its energy scales as $t^{-1/3}$ due to adiabatic losses. Therefore, it is valid (up to a logarithmic factor) for every density profile in which the shock accelerates before the breakout. The spherical power law is determined by two factors: adiabatic losses, and the fact that inner shells radiate at later times. Its value is weakly dependent on $n$.

The rise-time, which is also the time it takes the planar phase to begin is $t_0 \approx d_0 / v_0$ and the transition time between the planar and spherical power laws is $t_{\rm{s}} \approx R_*/v_0$, where $d_0$ is the breakout shell's initial width, and $v_0$ is the matter velocity at breakout. The subscript $'0'$ denotes the breakout shell mass (Lagrangian) coordinate. The bolometric luminosity at the breakout is $L_0 = E_0 / t_0$ where $E_0 \approx \rho_0v_0^24\pi R^2 d_0$ is the energy contained within the breakout shell at breakout, $\rho_0$ is the initial density at the breakout shell.

Assuming that the flux is well approximated by blackbody radiation, its temperature is determined at the last point from which enough photons can be generated to maintain thermal equilibrium. A measure of the thermal coupling is
\begin{equation}
\eta(m,t) \equiv \frac{n_{\rm{BB}}}{\dot{n}t_{\rm d}},
\end{equation}
where $n_{\rm{BB}}$ is the photons number density in the shell (assuming it is in thermal equilibrium), $\dot{n}$ is the rate of photons emission in the shell (proportional to the absorption opacity via Kirchhoff's law of thermal radiation) and $t_{\rm d}$ is the diffusion time. At all times, $\eta(m)$ is monotonically decreasing with $m$ measured from the edge of the star. Shells with $\eta > 1$ do not produce enough photons to maintain thermal equilibrium, while shells with $\eta <1$ do, therefore the observed temperature is the temperature of the outermost shell which satisfies $\eta=1$. If at the time of breakout this shell is inner to the breakout shell (also the point where the observed flux is determined), the breakout is out of thermal equilibrium and thermal equilibrium is obtained only during the spherical phase. If, on the other hand, this shell is outer to the breakout shell, thermal equilibrium is maintained at all times, starting from the breakout.

Most RSG explosions are found to be in thermal equilibrium from the breakout on, and the temperature scales with $t^{-\beta}$ where $\beta_{\rm p}=0.36$ and $\beta_{\rm{s}}=0.56$ (equation 31 in NS10). The observed temperature at the peak of bolometric luminosity is $T_{\rm obs,0} \approx T_{\rm{BB},0}\eta_0^{0.14}$, where $T_{\rm{BB},0} \approx (\rho_0v_0^2/a_{\rm{BB}})^{1/4}$ is the temperature of the breakout shell at the time of the breakout, $a_{\rm{BB}}$ is the radiation constant, and $\eta_0$ is the value of $\eta$ in the breakout shell at the breakout time. Both $T_{\rm{BB},0}$ and $\eta_0$ are mainly dependent on $v_0$ (equations 16,18 in NS10). Note that since the temperature is not necessarily determined at the breakout shell, the observed temperature is different from the breakout shell temperature.

NS10 also derived the dependency of the breakout shell parameters ($v_0$, $\rho_0$, $d_0$, $\eta_0$, etc.) on the progenitor properties, namely $R_*$, the progenitor's radius, $M_{\rm ej}$ the ejecta mass, and $E_{\rm exp}$ the explosion energy (see appendix A, equations A-6,A-7,A-10,A-14 in NS10)\footnote{It is worth to note that $d_0$ is fully dependent on $v_0$ and $\rho_0$, since $\tau_0 \simeq c/v_0$ where $\tau_0 \approx [1/(n+1)]\kappa_T\rho_0d_0$ is the optical depth of the breakout shell and $\kappa_T$ is Thomson scattering opacity.}. Substituting this dependency into the model to describe the emission, yields relations between the emission properties and the progenitor properties.

A similar analysis has been done by RW11, and \citet{Chevalier1992}, who obtained analytic estimates for the spherical phase only, and by \citet{Piro2010} who obtained estimates for both phases but for breakouts from white dwarf progenitors only. All the analytic models assumed similar progenitor structures, and obtained similar temporal power-law indices for the bolometric luminosity. The numerical coefficients of the light curve features vary by a factor of $2-4$ between the studies. The temporal evolution of the temperature is a bit different between NS10 and the other studies, since only NS10 considered the thermal coupling in detail. \citet{Sapir2011} performed numerical calculations of the light curve at planar geometry, for a progenitor of $n=1.5$, and found that the value of $L_0=E_0/t_0$, as predicted by NS10, should be multiplied by a factor of $2$. The rest of the estimates given above have not yet been checked numerically, especially for the spherical phase light curve, and using a more realistic progenitor structure.

\section{Numerical light curves of analytic progenitors}
\label{section:semi_analytic}

The analytic models described in section \ref{section:theory} are limited in the manner that they require rough assumptions on the dynamics of the problem and on the initial conditions. Specifically, approximation 6 is inaccurate for realistic progenitors, and progenitors with different stellar structures might yield different hydrodynamic evolution leading to different light curves. Due to assumption 7, the analytic models inherently ignore the hydrodynamics of the inner envelope, and their effect on the light curve. A correction is required at times late enough for the inner envelope to dominate the emission, which may take place before recombination has begun. Assumption 8 does not affect the bolometric luminosity, which is independent of the breakout shell structure, but could affect the evolution of the observed temperature. Finally, even within the limit of validity of the analytic models their predicted luminosity and observed temperature are only correct to within an order of magnitude.

When coming to obtain a better analytic approximation it is useful to separate properties of the light curve which are more general form properties that depend more strongly on the exact progenitor structure. Therefore, we do not start by studying numerically calculated progenitors, each with a different density profile. Instead, in this section we study the emission from a set of analytic progenitors, all of which have the same density distribution but different masses and radii. The definition of an analytic progenitor allows us also to easily study the effect of each parameter (e.g., $R_*$, $M_{\rm ej}$) on the light-curve independently, and for a wide range of values. The density profile of the analytic progenitors that we consider is more realistic than the single power-law assumed in analytic calculations. It is a smooth connection of two power-laws, one from the center of the star ($\rho \propto r^{-k}$) and one from the edge of the star ($\rho \propto (R_*-r)^n$). Thus, in this section we relax assumptions 6-8 and study the effect of various values of $R_*$,$M_{\rm ej}$ and $E_{\rm exp}$ on the light curve. Since the emission at the breakout and during the planar phase depends only on the conditions at the breakout shell we also find a relation between the breakout shell parameters and the observed light curve (and between the breakout parameters and $R_*$,$M_{\rm ej}$ and $E_{\rm exp}$).

 %In this section we relax assumptions 5-8 by defining an analytic, but more realistic progenitor profile, which still has a power-law form near the edge, and numerically calculating the radiation emitted by artificially exploding it. This way, the breakout shell structure and properties are calculated rather than assumed. Then, we provide numerical "calibration factors" to the analytic estimates, thus obtaining a more accurate prediction of the emitted light curve for given . The definition of an analytic progenitor allows us to easily study the effect of each parameter on the light-curve independently, and for a wide range of values. The effect of the progenitor profile on the emission is studied in section \ref{section:numerical}, by considering a large set of different realistic profiles.

In order to simulate the explosion, we wrote a 1D spherical geometry, two-temperatures Lagrangian computer program and used it to calculate the shock propagation and emitted radiation after the shock breakout. Gravitation is neglected because $GM_{\rm ej}^2/R_* \ll E_{\rm exp}$, and effects such as nucleosynthesis or energy deposition due to $^{56}\rm{Ni}$ decay are neglected since they do not affect the light curve at early times. Therefore, our code solves the radiation hydrodynamic equations alone, under the diffusion approximation, thus keeping assumption 1 and approximations 2-3. In appendix \ref{section:code}, we describe the code in detail and show its results for standard test cases.

For the equation of state (EOS) of the matter, we chose that of an ideal gas, with $\gamma=5/3$, suitable for mono-atomic gas, and $\mu=0.6$ which corresponds to a fully ionized mixture of hydrogen and helium with primordial ratios. Radiation is approximated as an ideal gas with $\gamma=4/3$. The diffusion opacity includes Thomson scattering term, $\kappa_T=0.34$ cm$^2$/g, which corresponds to fully ionized hydrogen and helium with primordial ratios, in addition to an analytic estimation \citep{Zeldovich1967} of the absorption opacity from hydrogen free-free and bound-free interactions (this term is usually smaller than $\kappa_T$). This absorption opacity is also used to calculate the observed temperature (by post-processing). The opacities  we use are appropriate as long as the hydrogen is completely ionized. Therefore, our solution is limited for temperatures higher than $T \approx 7500 ^\circ $K.

The observed temperature is calculated by post processing the hydrodynamic profiles. At each time, the observed temperature, $T_{\rm obs}$, is defined as the temperature at the outermost point of the ejecta which is in thermal equilibrium. NS10 denote this point "color shell", while in other works it is denoted "thermalization depth". This definition of the observed temperature is applicable as long as the luminosity shell is in thermal equilibrium (see conditions for that in sub-section \ref{subsection:thermal}). As discussed in section \ref{section:theory}, at the color shell $\eta=1$, which is equivalent to the condition $\tau_{\rm abs}\tau=1/3$, where $\tau_{\rm abs}(m)$ is the Planck mean absorption optical depth from coordinate $m$ to the observer and $\tau(m)$ is the total diffusion optical depth (dominated by scattering). Thus, for each time-step we find the point where $\tau_{\rm abs}\tau=1/3$, and define $T_{obs}$ as the temperature at this location \footnote{We verify that $T_{obs}$ is not very sensitive to the exact definition, e.g., choosing a different numerical value for the thermalization condition, such as $\tau_{\rm abs}\tau_{\rm{s}}=1$ yields around $5\%$ difference in the observed temperature.}.

By including hydrogen bound-free interactions, our work further expands the analytic work of NS10, which only included hydrogen free-free interactions for the temperature determination (approximation 5). The effect of bound-free opacity on the thermal coupling and observed temperature is discussed in appendix \ref{section:opacity}, which also includes a comparison between the opacity of pure hydrogen, hydrogen and helium with primordial ratios, and solar metallicity. We find that in the temperatures and densities of interest, helium and metal bound-free transitions are negligible compared to hydrogen free-free and bound-free transitions, and therefore our choice of opacity describes the thermal coupling well.

\begin{figure}
    \centering
    \includegraphics[width=\linewidth]{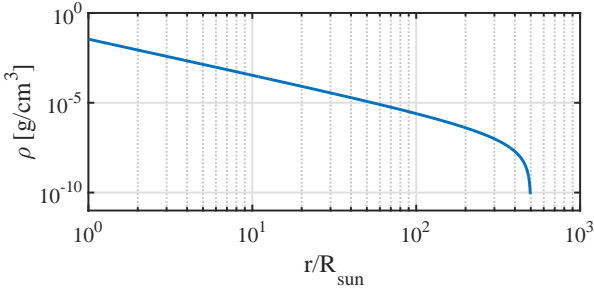}
    \caption{The analytic progenitor prototype density profile. The profile is composed of two power laws, one of the distance from the center and one of the distance from the edge (see text).}
    \label{fig:analytic_progenitor}
\end{figure}

As an analytic progenitor, we use a prototype with density profile $\rho(r) = Kr^{-k}(R_*-r)^n$. This density profile is a more realistic estimation of the envelope density, than the single power law approximation discussed at section \ref{section:theory}. We choose values of $k=2$ and $n=1.5$ which is a simple fit to the progenitors presented in \citet{Matzner1999} and to the ones calculated for this work (see section \ref{section:numerical}). The value of $K$ is determined by $M_{\rm ej}$. An example of the structure, for a progenitor with $M_{\rm ej}=15M_\odot$, $R_*=500R_\odot$ is shown in figure \ref{fig:analytic_progenitor}.

\subsection{The breakout shell parameters as a function of the progenitor properties}

\begin{figure}
    \centering
    \includegraphics[width=\linewidth]{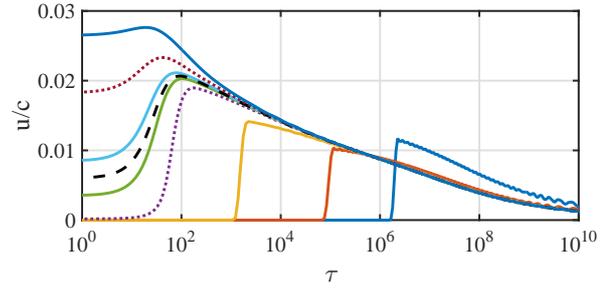}
    \caption{The velocity profile near the envelope edge at different times. $\tau$ is the diffusion optical depth to the observer. The breakout according to our definition is plotted in a black dashed line. The dotted lines mark the times where the luminosity rises from $10^{-4}$ of the peak (purple) to the peak (red).}
    \label{fig:analytic_breakout}
\end{figure}

The parameters of the breakout shell can be extracted from the code by looking at the hydrodynamic properties when $\tau \simeq c/v$. In order to be consistent, we defined the time of breakout at the point in which the velocity of the outermost mass element reached $1/3$ of the maximal velocity at that time. We found that this corresponds to $\tau_0=1.2c/v_0$, and to the point in which the luminosity reached $1/2$ of the maximal luminosity, for all the analytic progenitors. An example of the velocity profile near the envelope edge, for different times before and after the breakout is shown in figure \ref{fig:analytic_breakout}. At the snapshot time of the dotted purple line, the luminosity is practically zero, while at the snapshot time of the dotted brown line, the luminosity has already reached its peak. Thus, any definition of the breakout time must be between the two times that the snapshots depicted in the two dotted lines where taken. It can be seen that the maximal velocity changes by $20\%$ between the two dotted lines. Therefore, the definition of the breakout (the dashed black line) is robust, and the scaling relations obtained in this work are insensitive to the exact definition of the time of breakout.

Numerical calculations of the breakout were performed for progenitors of the same structure (See figure \ref{fig:analytic_progenitor}), with different radii, masses and explosion energies. The velocity, density and width of the breakout shell are found to scale within $1\%$ (the expected numerical error) as
\begin{subequations}
\label{eq:semi_analytic_hydro}
\begin{equation}
\label{eq:semi_analytic_hydro_rho}
\rho_0^A = 6.3 \cdot 10^{-10} \ \mathrm{g/cm^3} \ M_{15}^{0.67}R_{500}^{-1.64}E_{51}^{-0.31},
\end{equation}
\begin{equation}
v_0^A = 5000 \ \mathrm{km/s} \ M_{15}^{-0.44}R_{500}^{-0.24}E_{51}^{0.56},
\end{equation}
\begin{equation}
\label{eq:semi_analytic_hydro_d}
\frac{d_0^A}{R_*} = 2 \cdot 10^{-2} M_{15}^{-0.21}R_{500}^{0.9}E_{51}^{-0.25},
\end{equation}
\end{subequations}
where $M_{x}=M_{\rm ej}/xM_\odot$, $R_{x}=R_*/xM_\odot$ and $E_{x}=E_{\rm exp}/10^x$erg. The superscript $A$ notates the results for the analytic progenitors. We note that, as expected, 
\begin{equation}
\label{eq:semi_analytic_tau}
d_0 = \frac{2.5c}{\kappa_T\rho_0v_0}.
\end{equation}

The power law scaling relations are all similar to the ones defined in appendix A of NS10, but the numerical coefficients are different by factors ranging between $1.1$ and $3$. The breakout shell temperature is found to be
\begin{multline}
\label{eq:semi_analytic_Tbb}
T_{\rm{BB},0} = 1.2(\rho_0v_0^2/a_{\rm{BB}})^{1/4}\\
=4.5 \cdot 10^5 \ \mathrm{^\circ K} \ M_{15}^{-0.05}R_{500}^{-0.53}E_{51}^{0.2}.
\end{multline}

\subsection{Bolometric light curve and observed temperature as functions of the breakout properties}
\label{subsection:semi_analytic_lc}

Using the same progenitors, the scaling of the luminosity and observed temperature evolution with the breakout parameters can be found. From here on, we define the time that the bolometric luminosity peaks as $t=0$. We focus first on the emission after the peak and ignore light travel time (namely we ignore the difference in arrival time of photons that are emitted at the same time but from different locations on the expanding sphere). A typical light curve obtained using the simulation is shown at the top of figure \ref{fig:analytic_lc_calibrated}. It is best fit to a broken power law analytic formula with the form:

\begin{figure}
    \centering
    \includegraphics[width=\linewidth]{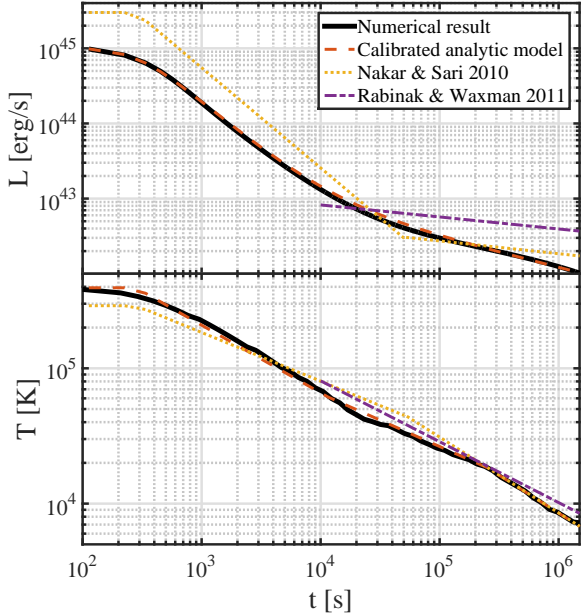}
    \caption{The bolometric luminosity and observed temperature of a typical analytic RSG progenitor. $t$ is measured from the peak of the bolometric luminosity. Light travel time is ignored. The numerical result (solid black line) is compared to the calibrated analytic model (dashed red line), and to the analytic models of NS10 (dotted yellow line) and RW11 (dotted-dashed purple line). The progenitor and explosion properties are $M_{\rm ej}=15M_\odot$, $R_*=500R_\odot$ and $E_{\rm exp}=10^{51}\rm{erg}$.}
    \label{fig:analytic_lc_calibrated}
\end{figure}

\begin{equation}
\label{eq:semi_analytic_L}
L_{\rm obs}(t) \simeq L_0
\begin{cases}
1 & t \ll t_0 \\
\left(\frac{t}{t_0}\right)^{-4/3} & t_0 \ll t \ll t_{\rm{s}} \\
\left(\frac{t_{\rm{s}}}{t_0}\right)^{-4/3}\left(\frac{t}{t_{\rm{s}}}\right)^{-0.35} & t_{\rm{s}} \ll t
\end{cases}.
\end{equation}

The best fit for the diffusion time at the breakout, planar to spherical transition time and maximal flux is:

\begin{subequations}
\label{eq:semi_analytic_lc}
\begin{equation}
\label{eq:semi_analytic_t0}
t_0 = d_0/5v_0,
\end{equation}
\begin{equation}
\label{eq:semi_analytic_ts}
t_{\rm{s}} = R_*/6v_0,
\end{equation}
\begin{equation}
L_0 \equiv E_0/t_0 = 1.7\rho_0v_0^32\pi R_*^2.
\end{equation}
\end{subequations}

The numerical coefficients for the transition times compared to the non-calibrated analytic prediction shows that the latter is indeed accurate only to within an order of magnitude. In order to improve the analytic approximation we use a smooth broken power law. At the transition between the planar and the spherical phases, the luminosity is well described by a sum of the planar and spherical luminosity, while at the transition between the breakout and the planar phase an harmonic sum of squares of the constant and planar luminosity is proper:

\begin{equation}
\label{eq:semi_analytic_L_transition}
L_{\rm obs}(t)=
\begin{cases}
\left(L_{t \ll t_0}^{-2} + L_{t_0 \ll t \ll t_{\rm{s}}}^{-2}\right)^{-0.5} & t \approx t_0 \\
L_{t_0 \ll t \ll t_{\rm{s}}} + L_{t_{\rm{s}} \ll t} & t \approx t_{\rm{s}}
\end{cases}.
\end{equation}

\begin{figure}
    \centering
    \includegraphics[width=\linewidth]{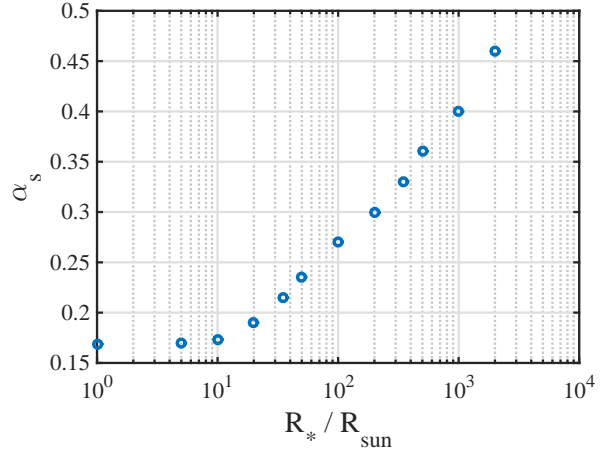}
    \caption{The temporal power law index for the spherical phase of the bolometric luminosity as calculated numerically for analytic progenitors with $M_{\rm ej}=15M_\odot$, $E_{\rm exp}=10^{51}\rm{erg}$ and varying $R_*$. Progenitors with radii $R_* < 10R_\odot$ yield an index value $0.17$ which is similar to the analytic model, but progenitors with $R_* \approx 500R_\odot$ (typical for RSG progenitors) yield a more rapid decrease.}
    \label{fig:analytic_alpha_R}
\end{figure}

As illustrated in figure \ref{fig:analytic_lc_calibrated}, this model fits the numerical calculation to within $5\%$ accuracy at all times. The planar phase luminosity power law of $\alpha_{\rm p} = 4/3$ is the one predicted by the analytic models (see section \ref{section:theory}). The spherical phase luminosity, however, is characterized by a decreasing power law $\alpha_{\rm{s}} = 0.3 - 0.4$, which is more rapid than the predicted value of $0.17$. The steeper decrease is due to the large radius of RSG progenitors, which yields a long planar - spherical transition time. For those progenitors, by the time the spherical phase has begun, the light curve is already dominated by inner parts of the envelope, where the density is not a pure power law and $(R_* - r) / R_* \sim 1$, so the analytic models fail to describe the emission accurately. Figure \ref{fig:analytic_alpha_R} shows the spherical temporal power law index as a function of $R_*$, obtained by calculations of progenitors with radii much smaller and larger than that of a typical RSG radius, $M_{\rm ej}=15M_\odot$ and $E_{\rm exp}=10^{51}\rm{erg}$. The spherical power in every calculation is obtained by best fitting the spherical emission to a power law until $t=15$ d. As seen in figure \ref{fig:analytic_alpha_R}, progenitors with $R_* < 10R_\odot$ yield a spherical light curve with a decay power law index of $0.17$ as expected, but for larger progenitors the power law index increases slowly with the radius. For the typical radii range of RSGs it is between $0.3$ and $0.4$.

Substituting equations \ref{eq:semi_analytic_hydro} into equations \ref{eq:semi_analytic_lc} yields direct relations between the progenitor properties and the light curve properties. The relations describe the light curve to within $5\%$. For comparison, results of the analytic models, as described in section \ref{section:theory} are also shown in figure \ref{fig:analytic_lc_calibrated}. Equations \ref{eq:semi_analytic_hydro}-\ref{eq:semi_analytic_L_transition}, which are basically a calibrated version of the analytic results, provide a much more accurate description of the light curve.

%When the breakout is in thermal equilibrium (for further discussion of the conditions for thermal equilibrium see sub-section \ref{subsection:thermal}), the observed temperature is determined at the last point in which enough photons can be generated to maintain thermal equilibrium. NS10 denote this point "color shell", while in other works it is denoted "thermalization depth". As discussed in section \ref{section:theory}, at the color shell $\eta=1$, though an equivalent definition for it is the point where $\tau_{\rm abs}\tau_{\rm{s}}=1/3$. Here, $\tau_{\rm{s}}(m)$ is the diffusion optical depth from coordinate $m$ to the observer, dominated by scattering processes, and $\tau_{\rm abs}(m)$ is the optical depth when taking into account only absorption-emission processes. A detailed analysis regarding the opacity appears in appendix \ref{section:opacity}.

%In this work, calculation of the observed temperature was done by post processing the hydrodynamic profiles at different times. For each time-step, the color shell and the observed temperature were found using the second definition\footnote{Choosing a different numerical value, such as $\tau_{\rm abs}\tau_{\rm{s}}=1$ yields around $5\%$ difference in the observed temperature.} $\tau_{\rm abs}\tau_{\rm{s}}=1/3$. 
The observed temperature which corresponds to the light curve calculation discussed above, is shown at the bottom of figure \ref{fig:analytic_lc_calibrated}, and compared with the non calibrated analytic models. The temperature evolution is somewhat different than the analytic predictions. While the prediction of NS10 is characterized by two power laws, one for the planar phase and one for the spherical phase, the temperature is better characterized by three power laws, corresponding to three phases:

\begin{multline}
\label{eq:semi_analytic_T}
T_{\rm obs}(t) \simeq T_{\rm{BB},0}\eta_0^{0.07} \times \\
\begin{cases}
1 & t < t_0 \\
\left(\frac{t}{t_0}\right)^{-0.45} & t_0 \leq t < t_{\rm{s}} \\
\left(\frac{t_{\rm{s}}}{t_0}\right)^{-0.45} \left(\frac{t}{t_{\rm{s}}}\right)^{-0.35} & t_{\rm{s}} \leq t < t_{\rm{c}} \\
\left(\frac{t_{\rm{s}}}{t_0}\right)^{-0.45} \left(\frac{t_{\rm{c}}}{t_{\rm{s}}}\right)^{-0.35} \left(\frac{t}{t_{\rm{c}}}\right)^{-0.6} & t_{\rm{c}} \leq t \\
\end{cases}.
\end{multline}

The temperature temporal evolution and the difference from NS10 predictions can be understood as follows. During the planar phase, the color shell is outer to the breakout shell (relative to the center of the star). Due to the planar nature of the evolution, matter - radiation coupling increases, and the color shell propagates outwards with time. Since the structure of the breakout shell is not well described by pure hydrodynamic models, NS10 roughly estimated it as a simple rarefaction wave (assumption 8 in section \ref{section:theory}). In the numerical calculation, we found a different evolution of the breakout shell structure, which origins in the fact that the initial density profile is not constant. As a result, the color shell propagates outwards faster than predicted, and the temperature drops more rapidly than predicted with $\beta_P=0.45$. By the transition to the spherical phase, side-way expansion becomes important and the envelope becomes more transparent. Therefore, the coupling decreases and the color shell propagates inwards. Nevertheless, the color shell is still outer than the breakout shell. NS10 neglected this phase, as they assumed that when the spherical phase begins, the color shell propagates very rapidly to a point inner to the breakout shell. We find here that this phase cannot be neglected and that during this phase $\beta_{S,1}=0.35$. Only when the thermalization point reaches the breakout shell, a third phase begins. At this phase, the temperature drops with $\beta_{S,2}=0.6$ as predicted by NS10 for the spherical phase, because the hydrodynamic evolution at locations inner than the breakout shell is well described by their model.

In addition to the different temporal behavior, the typical parameters of the observed temperature should be calibrated. We first find that $\eta_0$ is better estimated as

\begin{equation}
\label{eq:semi_analytic_eta}
\eta_0 
\approx 0.8\left(\frac{v_0}{10^4\mathrm{km/s}}\right)^\frac{15}{4}\left(\frac{\rho_0}{10^{-9}\mathrm{g/cm^3}}\right)^{-\frac{1}{8}}.
%\approx 3 \cdot 10^{-3} M_{15}^{-1.7}R_{500}^{-0.72}E_{51}^{2.12}
\end{equation}

This scaling is similar to equation 10 in NS10, only with a higher numerical factor\footnote{The lower coupling, reflected by the higher numerical factor, is a net result of the inclusion of bound-free transitions to the opacity, which increases the coupling by a factor of $4$ at breakout temperatures, and the higher breakout shell temperature (see equation \ref{eq:semi_analytic_Tbb}) and shorter diffusion time (see equation \ref{eq:semi_analytic_t0}) both decrease the coupling.}. It is obtained by examining several calculations with $T_{\rm{BB},0} \approx T_{\rm obs}$, and demanding $\eta_0=1$ for breakouts in which $T_{\rm{BB},0}=T_{\rm obs}$, i.e., the breakout shell is the color shell. Besides the different value of $\eta_0$, the initial observed temperature, given in equation \ref{eq:semi_analytic_T} scales as $\eta_0^{0.07}$ instead of $\eta_0^{0.14}$ as predicted. This is also due to the differences in the hydrodynamic structure of the breakout shell. 

The diffusion time at the breakout and the planar to spherical transition time are the same as in the luminosity analysis, and are given in equations \ref{eq:semi_analytic_lc}. A scaling relation for the third phase transition time can be obtained by observing equation 18 at NS10 and noting that for the breakout shell
\begin{equation}
\eta(m_0,t)=\eta_0\left(\frac{t_{\rm{s}}}{t_0}\right)^{-1/6}\left(\frac{t}{t_{\rm{s}}}\right)^\frac{42n+49}{12(1.19n+1)}.
\end{equation}
Since the breakout shell is also the color shell when $\eta=1$, we deduce the transition time is
\begin{equation}
\label{eq:semi_analytic_tc_general}
t_{\rm{c}}=6.5t_{\rm{s}}\left(\frac{1}{\eta_0}\right)^\frac{12(1.19n+1)}{42n+49}\left(\frac{t_{\rm{s}}}{t_0}\right)^\frac{2(1.19n+1)}{(42n+49)}.
\end{equation}
For RSG progenitors ($n=1.5$) we obtain
\begin{equation}
\label{eq:semi_analytic_tc}
t_{\rm{c}}=6.5t_{\rm{s}}\eta_0^{-0.3}\left(\frac{t_{\rm{s}}}{t_0}\right)^{0.05}.
\end{equation}
The numerical calibration factor of $6.5$ is chosen to best fit the numerical simulations. This model fits the numerical temperature well, as demonstrated in figure \ref{fig:analytic_lc_calibrated}.

\subsection{The breakout pulse}
\label{subsection:semi_analytic_rise}

\begin{figure}
    \centering
    \includegraphics[width=\linewidth]{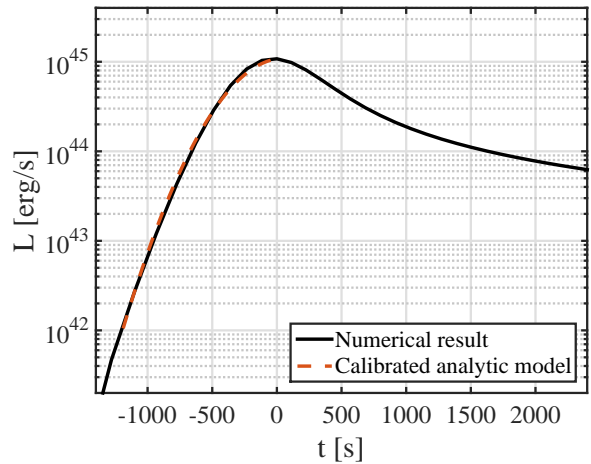}
    \caption{The bolometric luminosity of a typical analytic RSG progenitor during the breakout pulse. The numerical result (solid black line) is compared to the calibrated analytic model (dashed red line), described in equation \ref{eq:semi_analytic_rise}. The progenitor and explosion properties are $M_{\rm ej}=15M_\odot$, $R_*=500R_\odot$ and $E_{\rm exp}=10^{51}\rm{erg}$.}
    \label{fig:analytic_rise}
\end{figure}

During the breakout, the typical timescale is $t_0$, which is also the diffusion time at breakout (equation \ref{eq:semi_analytic_lc}). At early stages of the rise (before breakout), emission is due to photons diffusing ahead of the shock when the shock is far from the edge. The fraction of photons that diffuse a length $x$ ahead of the shock is $e^{-x^2}$, and since the shock distance from the edge at each time is approximately $x=-v_0t$ ($t$ is negative), the early rise should be dominated by a $e^{-t^2}$ term. Around the breakout (i.e. $\tau \simeq c/v$), however, the energy scales as $e^{-x}$ \citep{Weaver1976}. We therefore follow a method similar to the one discussed in \citet{Sapir2011}, and approximate the light curve during the rise ($t<0$) as
\begin{equation}
\label{eq:semi_analytic_rise}
L(t)=L_0e^{-a(t/t0)^2-b(t/t0)},
\end{equation}
where $L_0$,$t_0$ are defined in equation \ref{eq:semi_analytic_lc}. The best fit to the rise is obtained for $a=0.35$, $b=-0.15$ and is shown in figure \ref{fig:analytic_rise}.

The observed temperature during the breakout changes by less than $5\%$ from the time in which $L(t)=0.1L_{\rm peak}$ to the peak time, and therefore it can be approximated as a constant.

\section{Numerical light curves of numerical progenitors}
\label{section:numerical}
In this section we study the light curves generated by explosions of more realistic progenitors, whose structure is calculated numerically using a stellar evolution code. Since the progenitor structure depends on various initial parameters such as mass, rotation and metallicity, and on unknown fudge factors used by the code, such as a mixing length coefficient, we have calculated a large set of progenitors and calculated the light curves that they generate upon explosion. The numerical progenitors have unique profiles, and specifically their density near the edge is not well characterized by a single power law with index $n$.

\begin{figure}
    \centering
    \includegraphics[width=\linewidth]{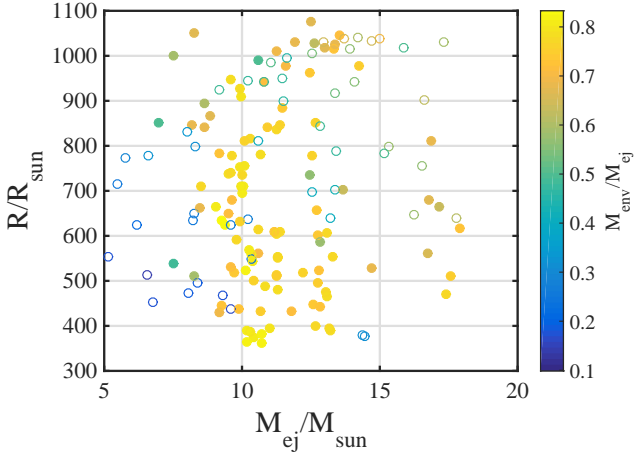}
    \caption{The properties of the stars that were numerically calculated using MESA, and classified as realistic type II-P/II-L progenitors. The different colors represent different values of the ratio $M_{\rm env}/M_{\rm ej}$, where $M_{\rm env}$ is the envelope mass and $M_{\rm ej}$ is the ejecta mass. The filled circles represent progenitors with $M_{\rm ZAMS} \leq 20M_\odot$ which are more common.}
    \label{fig:numeric_progenitors}
\end{figure}

The stellar evolution of the progenitor models was followed using the publicly available package MESA version 6596 \citep{2011ApJS..192....3P,2013ApJS..208....4P,2015ApJS..220...15P}. To produce a wide range of progenitors, we varied the zero age main sequence (ZAMS) mass between $[10, 50] M_{\sun}$, the metallicity between $[2 \times 10^{-5}, 2 \times 10^{-2}]$, the mixing length parameter between $[1.5,5]$, and the initial rotation rate between $[0,0.8]$ of the breakup rotation rate. In all models, mass loss was determined according to the "Dutch" recipe in MESA, combining the rates from \cite{2009A&A...497..255G, 1990A&A...231..134N, 2000A&A...360..227N, 2001A&A...369..574V}, with a coefficient $\eta=1$, the convection was according to the Ledoux criterion, with a semi-convection efficiency parameter $\alpha_{sc}=0.1$ \citep[eq. 12]{2013ApJS..208....4P}, and exponential overshoot with parameter $f=0.008$ \citep[eq. 2]{2011ApJS..192....3P}.

A total of $219$ progenitors were calculated. Out of them, we chose $124$ who are more realistic type II-P/II-L SN candidates, by choosing progenitors with $M_{\rm env} \geq 4M_\odot$ and $R_* \geq 100R_\odot$. The stars that were cut do not have large enough radii and mass to emit detectable cooling envelope radiation for several weeks. In order to simulate the explosion, we cut out the $^{28}\rm{Si}$ core, and referred to everything outside that core as ejecta. Then, we planted the explosion energy as thermal energy in the $5$ innermost cells of the ejecta. The properties of the different progenitors are specified in figure \ref{fig:numeric_progenitors}, where the envelope begins at boundary between the helium shell and the hydrogen shell, where a sharp density drop exists. Full data is given in appendix \ref{section:numerical_progenitors}.

\subsection{Bolometric light curve and observed temperature as functions of the breakout properties}
\label{subsection:numeric_lc}

\begin{figure}
    \centering
    \includegraphics[width=\linewidth]{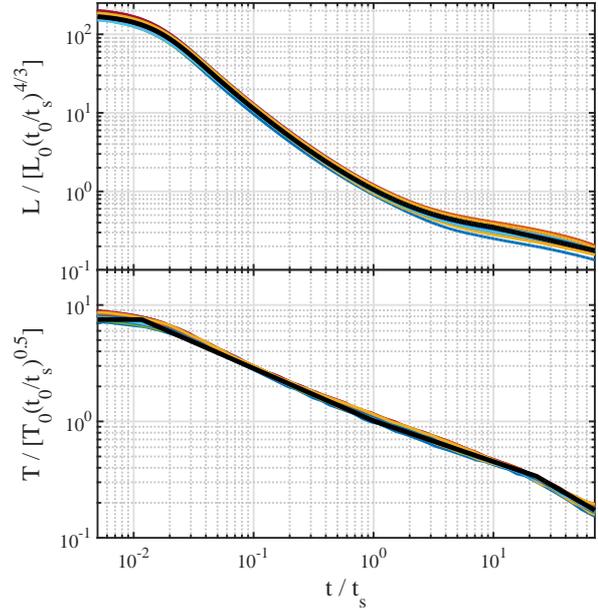}
    \caption{Bolometric luminosity and observed temperature obtained numerically for 20 numerical RSG progenitors from our sample (light travel time ignored). The time for each light curve is normalized by $t_{\rm{s}}$ as obtained by the calibrated analytic model. $L$ and $T$ are normalized by the their value at $t_{\rm{s}}$, as predicted by the analytic model. Also shown is a normalized light curve of the calibrated analytic model (black thick line). This figure shows that all progenitors produce rather similar light curve shapes, with a smaller spread before $t_{\rm{s}}$ than after $t_{\rm{s}}$. It also illustrates that the calibrated analytic model provides a good fit to the numerical one.}
    \label{fig:numeric_all_lc}
\end{figure}

In figure \ref{fig:numeric_all_lc}, typical normalized light curves from $20$ different progenitors are presented. The time axis is normalized to the model spherical time (equation \ref{eq:semi_analytic_ts}), and the bolometric luminosity and observed temperature are normalized to the values given by the model at $t=t_{\rm{s}}$, (equations \ref{eq:semi_analytic_L}, \ref{eq:semi_analytic_T} respectively). Namely, the normalization is done based on the parameters of the breakout shell. Figure \ref{fig:numeric_all_lc} also shows our analytic model (equations \ref{eq:semi_analytic_L}-\ref{eq:semi_analytic_eta} and \ref{eq:semi_analytic_tc}), normalized similar to the numerical light curves (black thick line). The figure shows that all the light curves are similar, and that the calibrated analytic model (which relates breakout parameters to the light curve) fits the numerically calculated progenitors as well as the analytic ones. In fact, the relations between the luminosity and temperature evolution and the properties of the breakout shell, that were specified in equations \ref{eq:semi_analytic_L}-\ref{eq:semi_analytic_eta} and \ref{eq:semi_analytic_tc}, fit the emission of the numerical progenitors without further adjustments. It fits the luminosity to within $15\%$ and the temperature to within $10\%$ until well within the spherical phase. At $t \gtrsim 4t_{\rm{s}}$ there is a slightly larger spread in the luminosity evolution of different progenitors, since the spherical phase luminosity power-law index $\alpha_{\rm s}$, is somewhat dependent on the inner progenitor structure, and because larger progenitors yield a more rapid luminosity decrease (as described in section \ref{section:semi_analytic}). The luminosity fit is then accurate to within $30\%$.

The similar light curve evolution of different progenitors during the planar phase is expected and it is a result of the light curve being mostly determined at the breakout shell. The spherical phase however probes inner layers of the progenitor and therefore the observed similarity is less obvious. The similar emission of different progenitors during the spherical phase implies that the early light curve depends weakly on the density profile, as suggested by the weak dependence on $n$ in analytic progenitors. We therefore expect the model based on the breakout properties to be applicable for a wide range of structures including such that are significantly different than those in our sample.

\begin{figure}
    \centering
    \includegraphics[width=\linewidth]{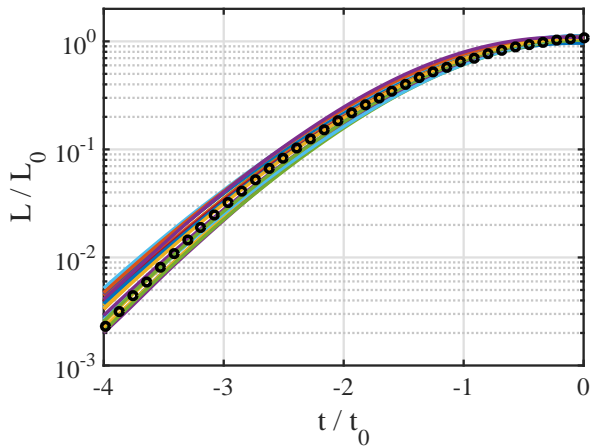}
    \caption{Bolometric light curves during the breakout pulse, of 20 numerical RSG progenitors. The luminosity is normalized to $L_0$ and the time is normalized to $t_0$ (equation \ref{eq:semi_analytic_t0}). The analytic model for the rise, given in equation \ref{eq:semi_analytic_rise} is marked in black circles.}
    \label{fig:numeric_all_rise}
\end{figure}

The breakout pulse is also well approximated by the analytic model specified in equation \ref{eq:semi_analytic_rise}. The numerical light curves and analytic model are shown in figure \ref{fig:numeric_all_rise}. During the last decade of rise, the analytic model fits the numerical results better than $30\%$, while earlier it fits to a factor of $2$.

\subsection{The observed spectrum}
\label{subsection:numeric_spectrum}

Earlier analytic works, as well as some numerical studies, assumed the observed spectrum was a blackbody with temperature $T_{\rm obs}$ (approximation 4 in section \ref{section:theory}). However, this is not exactly true. At high frequencies ($h\nu>3 k_B T$, where $k_B$ is Boltzmann constant and $h$ is Planck constant) the emission is suppressed due to line blanketing, while at low frequencies ($< k_B T$), namely the Rayleigh-Jeans regime, a deviation is expected even when line emission and absorption is neglected. Deviations from blackbody spectrum at the Rayleigh-Jeans regime are especially important during very early times (breakout and planar phase), when the temperature is $\approx 10^{5 ~\circ}$K and most of the radiation is emitted at frequencies higher than the optical/UV bands. During these times, even a small deviation from a pure blackbody can significantly affect the observed light curve. As we explain below, such deviation is expected, and also seen in simulation results where radiation transfer is solved more accurately than our code (e.g., \citealt{Tominaga2011}). Below we derive an analytic approximation to the observed spectrum in the Rayleigh-Jeans regime and compare it to the results of \cite{Tominaga2011}.

\begin{figure}
    \centering
    \includegraphics[width=\linewidth]{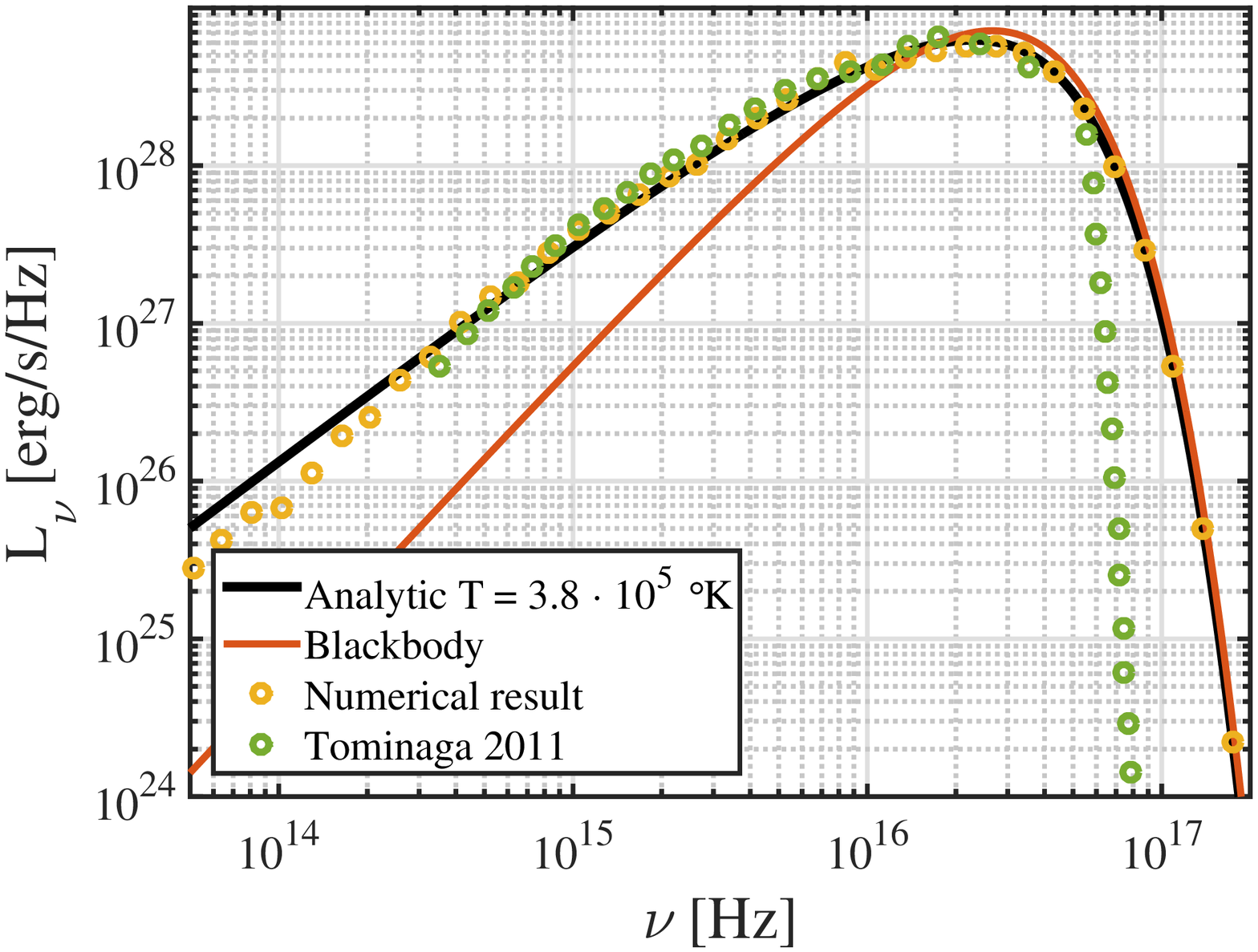}
    \caption{The spectrum emitted from the same progenitor as in figure \ref{fig:numeric_lc_calibrated} at $t \approx 300$s after the breakout, when $T_{\rm obs}=3.8 \cdot 10^5 \ \mathrm{^\circ K}$. The analytic model (solid blue line) fits both our numerical result (yellow circles) and the result of \citet{Tominaga2011} (green circles, see text) to within $20\%$ at frequencies belowe the spectral peak. At frequencies much above the peak the spectrum of \citet{Tominaga2011} falls faster as they account for line blanketing while we do not. A blackbody spectrum at the same temperature is depicted for comparison (solid red line).}
    \label{fig:numeric_spectrum_nobb}
\end{figure}

The color shell (a.k.a. thermalization depth) is the outermost point where a significant number of photons with $h\nu \sim k_B T_e$ is created (hence $\eta=1$). Here, $T_e$ is the local electron temperature. It is also the outermost point where such photons are absorbed (hence $\tau_{\rm abs}\tau=1/3$). Outer to this point, the energy flux is dominated by these photons and therefore the color temperature of the radiation is constant (and equals $T_{obs}$). However, at lower frequencies the spectrum continues to change, since photon opacity is frequency dependent, and so is the absorption optical depth to the observer, denoted here as $\tau_{\rm \nu,abs}$ (not to be confused with $\tau_{\rm abs}$, which is the Planck mean optical depth). When free-free and bound-free processes dominate the absorption, $\tau_{\rm \nu,abs}$ is larger for lower frequency photons. As a result, the number of photons with $h\nu<k_B T_{obs}$ is set outer to the color shell, at the point where $\tau_{\rm \nu,abs}\tau \approx 1/3$ and the electron temperature, $T_e$, is lower than $T_{\rm obs}$. We use this criterion and post process the observed spectrum from our numerical hydrodynamic profiles. An example of such spectrum is depicted in figure \ref{fig:numeric_spectrum_nobb}. \footnote{We note that although our code does allow for different photon and electron temperatures, the calculation of $T_e$ in regions where the radiation spectrum is not a blackbody is not fully accurate as the heating term in equation \ref{eq:de_dt} implicitly assumes a blackbody spectrum. Nevertheless, since the drop in electron temperature at lower optical depth depends mostly on the drop in the radiation energy density, which is accounted for in the code, it does provide a reasonable approximation of $T_e$.}

%Here, we relax this approximation, only assuming that the radiation is in thermal equilibrium where $\eta<1$. We still neglect line emission and absorption as well as line blanketing, which is a reasonable approximation near the peak of the spectrum and in the Rayleigh Jeans regime. Note that although thermal equilibrium is assumed everywhere, the observed spectrum is not blackbody, since the luminosity in different wavelength is determined at different points of the expanding envelope. Here we derive the observed spectrum from the bolometric luminosity and observed temperature.
%Using the numerical results, a spectrum can be calculated by considering the point where $\tau_s \tau_{\rm abs}(\nu)=1/3$. Here, $\nu$ is the photon frequency, and $\tau_{\rm abs}(\nu)$ is the absorption opacity (not averaged), which is highly dependent on $\nu$. An example numerical spectrum is depicted in figure \ref{fig:numeric_spectrum_nobb}. 

In order to obtain an analytic approximation for the spectrum, we approximate the density profile during the expansion as a power-law, 
\begin{equation}
\label{eq:rho_tau}
\rho(\tau) \propto \tau^k.
\end{equation}
In most of our progenitors, $k \approx 1$ around the breakout and later (the analytic solution yields $k = 11/12$ during the spherical phase). Outer to the luminosity shell, the luminosity is constant and it is $L \propto r^2 U_r/\tau$ where $U_r$ is the radiation energy density. During the breakout and the planar phase $r$ is roughly constant, and during the spherical phase the sharp density gradient dictates that $r$ varies slowly with $\tau$. Therefore we can approximate $U_r \propto \tau$. By assuming $U_r=a_{BB}T_e^4$ we obtain
\begin{equation}
\label{eq:T_tau}
T_e(\tau) = T_{\rm obs}\left(\frac{\tau}{\tau_{\rm obs}}\right)^{1/4},
\end{equation}
where $\tau_{\rm obs}$ [$\tau_{\rm abs,obs}$] is the diffusion [absoprion] optical depth where $T=T_{\rm obs}$. The absorption opacity (bound-free and free-free) is approximately proportional to 
\begin{equation}
\label{eq:kappa_tau}
\kappa_{\rm \nu,abs}(r) \propto \rho T_e^{-0.5}\nu^{-3}(1-e^{-h\nu/k_BT_e}).
\end{equation}
Substituting equations \ref{eq:rho_tau} - \ref{eq:T_tau} into equation \ref{eq:kappa_tau} we obtain the following expression for the absorption opacity at the Rayleigh-Jeans regime ($h\nu \ll k_BT_{\rm obs}$):
\begin{equation}\label{tau_nu}
\tau_{\rm \nu,abs}(r) = \tau_{\rm abs,obs}
\left(\frac{h\nu}{3k_BT_{\rm obs}}\right)^{-2} \left(\frac{\tau}{\tau_{\rm obs}}\right)^{\frac{15k+8}{8(k-1)}}.
\end{equation}
We define $\tau_{\rm col}(\nu)$ as the optical depth at the point where $\tau_{\rm \nu,abs} \tau_{\rm col}(\nu)=1/3$. Equation \ref{tau_nu} dictates then
\begin{equation}
\label{eq:tau_col}
\tau_{\rm col}(\nu) = \tau_{\rm obs}
\left(\frac{h\nu}{3k_BT_{\rm obs}}\right)^{\frac{16k}{13k+8}}.
\end{equation}
Since $k \approx 1$ in most cases, the power of $\nu$ is approximately $16/21$. Substituting equation \ref{eq:tau_col} into equation \ref{eq:T_tau} we find that the electron temperature observed at each frequency is
\begin{equation}
\label{eq:T_col}
T_{\rm col}(\nu) \approx
T_{\rm obs}\left(\frac{h\nu}{3k_BT_{\rm obs}}\right)^{0.2} ~~~;~~~ h\nu < k_BT_{\rm obs}
\end{equation}
for the Rayleigh-Jeans regime. Since the luminosity is constant outer to the luminosity shell, the frequency dependence of $T_{\rm col}$ implies that the modified Rayleigh-Jeans spectrum is $L_\nu \propto \nu^{1.4}$. At high frequencies ($h\nu > 3k_BT_{\rm obs}$) we assume that the thermalization depth is $\tau_{\rm obs}$, meaning $T_{\rm col}(\nu) =  T_{\rm obs}$. By comparing the analytic model to numerical results we find that the spectrum, including the peak, is best fit as a harmonic sum
\begin{equation}
T_{\rm col}(\nu) = T_{\rm obs} \left(1+\left(\frac{h\nu}{3k_BT_{\rm obs}}\right)^{-0.2*m}\right)^{-1/m}
\end{equation}
with $m > 10$ (sharp transition at $h\nu = 3k_BT_{\rm obs}$). The luminosity at each wavelength is then given by 
\begin{multline}
\label{eq:spectrum_nobb}
L_\nu = 0.9 L \cdot \frac{15}{\pi^4}\left(\frac{h}{k_BT_{\rm col}(\nu)}\right)^4\nu^3 \cdot \\ \left(e^{\left(\frac{h\nu}{k_BT_{\rm col}}\right)}-1\right)^{-1},
\end{multline}
The factor of $0.9$ is obtained by demanding the integral of the spectrum over $\nu$ to equal to $L$. Figure \ref{fig:numeric_spectrum_nobb} depicts a comparison between eauation \ref{eq:spectrum_nobb} and a spectrum calculated numerically by post-processing the hydrodynamic profiles obtained short time after a breakout (in this case $T_{\rm obs}=3.8 \cdot 10^5 \ \mathrm{^\circ K}$ and $L=1.6 \cdot 10^{45}$ erg/s). Generally, we find that for the optical and UV bands the analytic model deviates by less than $25\%$ from the numerical results at all times.

While deriving equation \ref{eq:spectrum_nobb} we assumed that $T_{\rm col}(\nu)$ is set in a diffusion optically thick region, namely $\tau_{\rm col}(\nu) > 1$. Therefore, equation \ref{eq:tau_col} implies that equation \ref{eq:spectrum_nobb} is valid only for frequencies higher than a critical frequency which satisfies $h\nu \approx \tau_{\rm obs}^{-1.3} (3k_BT_{\rm obs})$. Below this frequency the spectrum it better described by a blackbody (i.e., $L_\nu \propto \nu^2$). For the progenitors and explosion energies we explored this critical frequency is typically around the optical bands.

An additional condition that must be satisfied for our analytic approximation to be valid is that the luminosity shell is {\it deeply} within thermal equilibrium, i.e., $\eta(\tau=c/v) \ll 1$. The reason is that when $\eta(\tau=c/v) \approx 1$ the luminosity shell is also the color shell and the electrons in this shell have just enough time to cool down (by emitting photons) so $U_r=a T_e^4$. Therefore, outer shells (with $\tau<c/v$) do not have enough time to cool down and our approximation of $U_r=a T_e^4$ is not valid. In fact when  $\eta(\tau=c/v) \sim 1$ a blackbody is probably a better approximation for the observed spectrum. This condition, $\eta(\tau=c/v) \ll 1$, is typically satisfied for RSG explosion as discussed in section \ref{subsection:thermal}.

In order to test our analytic (and numerical) spectra we compare our results to a spectrum presented in \citet{Tominaga2011}. They find the spectrum using the numerical code STELLA \citep{Blinnikov1998}, which uses a multi-group radiative transfer and does not need to employ many of the approximations we use here in order to derive the observed spectrum. Figure \ref{fig:numeric_spectrum_nobb} depicts, in addition to our analytic and numerical spectra, a spectrum taken from figure 2a ($t=0$) in \citet{Tominaga2011}. They calculated this spectrum at the breakout of an explosion with $E_{\rm exp}=10^{51}$erg, $R_*=795R_\odot$ and $M_{\rm ej}=16.8M_\odot$. Our spectra are taken from an explosion with $E_{\rm exp}=10^{51}$erg, $R_*=624 R_\odot$ and $M_{\rm ej}=9.3 M_\odot$. Therefore $L$ and $T_{obs}$ are expected to be slightly different. In figure \ref{fig:numeric_spectrum_nobb} we multiply their luminosity by a factor of $1.5$ and divide our breakout temperature by $1.15$, so the peak of their spectrum coincides with ours. The comparison shows a very good agreement of the spectral shape in the Rayleigh-Jeans regime and near the peak of $L_\nu$. Above the peak, their spectrum falls faster than ours since their code includes line blanketing while we neglect it.

%a comparison between the analytic model and a specific numericalA comparison to the work of \citet{Tominaga2011} is also included. The numerical result therein (figure 2) was obtained by exploding a progenitor with $R_*=795R_\odot$, $M_{\rm ej}=16.8M_\odot$ and $E_{\rm exp}=10^{51}$erg. Their code includes multigroup diffusion and nucleosynthesis. In order to account for the different progenitor parameters, we multiplied their luminosity by a factor of $1.5$ and chose the time where our temperature is $0.85$ of its value at breakout. The numerical factors of $1.5$ and $0.85$ are yielded from the scaling equations \ref{eq:final_L} and \ref{eq:final_T} accordingly. As can be seen, the analytic model fits both calculations very well for all the Rayleigh-Jeans regime. At the exponential fall, the result of \citet{Tominaga2011} falls faster, since line absorption is significant. 

\subsection{The effect of light travel time}
\label{subsection:ltt}

\begin{figure}
    \centering
    \includegraphics[width=\linewidth]{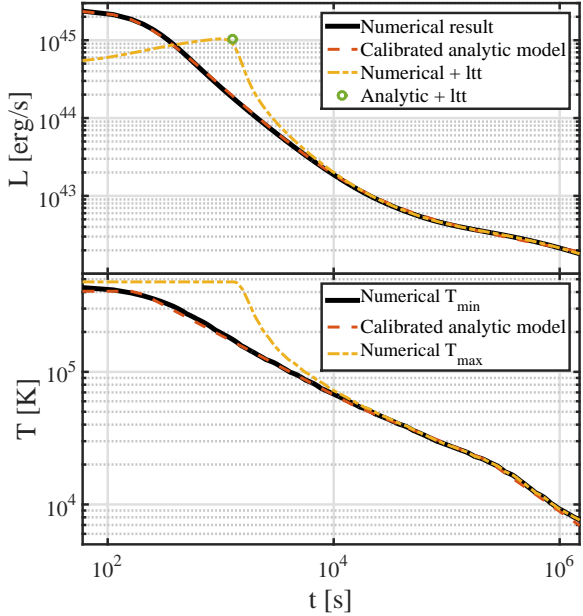}
    \caption{Top: the bolometric luminosity of a typical numerical RSG progenitor after the peak. Shown are the numerical result (solid black line) and the analytic model (dashed red line) where light travel time effects are neglected, and the same numerical result, with light travel time effects included (dotted-dashed yellow line). The analytic estimation to the peak with light travel time (equation \ref{eq:numeric_ltt_L}) is marked by a green circle. Bottom: the observed temperature of the same progenitor after the peak. Light travel time causes the observer to see a range of temperatures. $T_{\rm min}$ (solid black line) is the line-of-sight temperature (i.e., the same as in the model where light travel time is ignored), while $T_{\rm max}$ (solid yellow line) is calculated using equation \ref{eq:T_max}. The analytic model without light travel time is plotted in dashed red line. The progenitor and explosion properties are $M_{\rm ej}=9.3M_\odot$, $R_*=624R_\odot$ and $E_{\rm exp}=10^{51}\rm{erg}$. For this progenitor, $t_0=180\rm{s}$ and $t_{\rm Rc}=1450\rm{s}$.}
    \label{fig:numeric_lc_calibrated}
\end{figure}

The model described in sub-sections \ref{subsection:semi_analytic_lc}  \& \ref{subsection:numeric_lc} neglects light travel time, though it must be considered at times where $t \lesssim R_*/c$. Radiation emitted at small angles relative to the line which connects the source to the observer, travels a shorter distance to the observer and is thus detected earlier than radiation emitted at large angles. This yields a smearing of each point in the source frame light curve as a rectangular pulse of width $t_{\rm Rc}=R_*/c$ and height $L(t)/t_{\rm Rc}$ where $L(t)$ is the luminosity in the source frame \citep{Katz2012}. The luminosity observed at each time due to this effect is

\begin{equation}
\label{eq:ltt_general}
L_{\rm ltt}(t) = \frac{1}{t_{\rm Rc}}\int_{t-t_{\rm Rc}}^{t} L(t') dt'.
\end{equation}

For most RSG progenitors $t_{\rm Rc} \ll t_{\rm{s}}$, as demonstrated in section \ref{section:final}. Therefore, light travel time is important only during the breakout pulse and early in the planar phase. The exact observed luminosity can be obtained from the model of the source frame lumnostiy by accounting for the effect numerically (using equation \ref{eq:ltt_general}). For the convenience of the reader, we present a simple analytic estimate of the peak bolometric luminosity, including light travel time, and the time of the peak when light travel time is considered, relative to the peak when it is not.

\begin{figure}
    \centering
    \includegraphics[width=\linewidth]{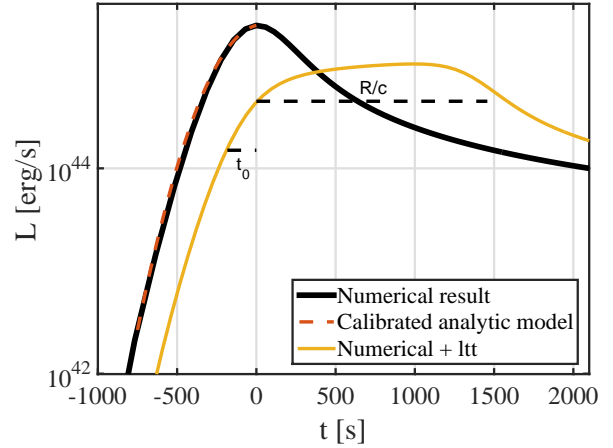}
    \caption{The bolometric luminosity of the same progenitor from figure \ref{fig:numeric_lc_calibrated} during the breakout pulse. Both the numerical result (solid black line) and the analytic model (dashed red line) neglect light travel time. The same numerical result, after accounting for light travel time (solid yellow line) is composed of an exponential rise at early times, a slower rise from $t=0$ to $t=t_{\rm Rc}$ and a rapid decline thereafter.}
    \label{fig:numeric_rise}
\end{figure}

Since the rise is much faster than the fall, only emission from the last $t_0$ before the peak contributes to the peak luminosity, and the peak time is approximately $t_{\rm peak,ltt}=t_{\rm Rc}-t_0$, where again $t=0$ is defined as the peak of bolometric emission without light travel time. The peak luminosity, light travel time included, is then given by
\begin{equation}
L_{0,ltt} = L_0\frac{t_0}{t_{\rm Rc}}+\frac{1}{t_{\rm Rc}}\int_{t=0}^{t_{\rm Rc}-t_0} L(t') dt',
\end{equation}
where $L(t')$ is given by equation \ref{eq:semi_analytic_L}. Integration (during the planar phase) yields
\begin{equation}
\label{eq:numeric_ltt_L}
L_{0,ltt} = L_0\frac{3t_0}{t_{\rm Rc}}f,
\end{equation}
where $f$ is a numerical factor which equals
\begin{equation}
\label{eq:numeric_ltt_f}
f = \frac{5}{3}-\left( \frac{t_{\rm Rc}}{t_0} - 1 \right) ^{-1/3}.
\end{equation}

The value of $f$ varies between $0.8$ and $1.2$ for most progenitors. Figure \ref{fig:numeric_lc_calibrated} shows the calculated emission for a typical numerical progenitor, after the peak, with and without light travel time (in log-log scale). In addition are shown the analytic model (without light travel time) and the analytic estimate for the peak luminosity and peak time including light travel time. The calculated light curve before and after the peak is shown in figure \ref{fig:numeric_rise} (in semi-log scale). During the early rise, the shape of the pulse is weakly affected by light travel time because of the rapid increase. Therefore, the smeared emission rises during a typical timescale $t_0$. Then, it rises slowly (changes by less than a factor of $2$) from $t=0$ to $t_{\rm peak,ltt}$, reaches the estimated peak value $L_{0,ltt}$ and falls faster than the planar power law during a typical timescale $t_0$ until it coincides with the source frame light curve.

\begin{figure}
    \centering
    \includegraphics[width=\linewidth]{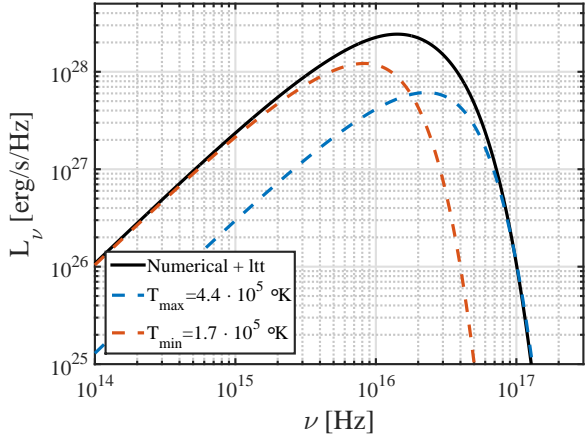}
    \caption{The spectrum emitted from the same progenitor as in figure \ref{fig:numeric_lc_calibrated}, at $t=t_{\rm Rc}$. The observed spectrum (solid black line) accounts for light travel time and is therefore non-thermal. The relative contribution of $T_{\rm max}$ (dashed red line) and $T_{\rm min}$ (dashed orange line) is also shown.}
    \label{fig:numeric_spectrum}
\end{figure}

Due to light travel time, the spectrum also changes. As the source frame temperature drops with time, low frequency photons arrive at the observer from small angles with respect to the line of sight at the same time that high frequency photons arrive from large angles. During the breakout and the exponential rise, only small angles contribute to the emission, so the spectrum is similar to the breakout spectrum at the source frame. From around the peak at the source frame, the spectrum is composed of three different regions: A modified Rayleigh-Jeans ($\nu^{1.4}$) radiance with the temperature at angle $\theta=0$, which we denote $T_{\rm min}$, a relatively constant radiance to the temperature at the largest angle that contributes to the emission $T_{\rm max}$, and an exponential fall. The calculated spectrum at $t=t_{\rm Rc}$ is shown in figure \ref{fig:numeric_spectrum}, and compared with the relative contribution of $T_{\rm min}$ and $T_{\rm max}$.

The bottom of figure \ref{fig:numeric_lc_calibrated} depicts the observed temperature without light travel time, which is also $T_{\rm min}$ when light travel time is considered, the analytic model, and $T_{\rm max}$, which is defined as\footnote{At times earlier than $t_{\rm Rc}-t_0$, $T_{\rm max}$ is not determined at $\theta=90^\circ$, since the emission at this angle is negligible compared to the emission from angles where light from the breakout has already reached the observer.}

\begin{equation}
\label{eq:T_max}
T_{\rm max} = T_{\rm obs}(\mathrm{max}(t-t_{\rm Rc},0)).
\end{equation}

\subsection{The breakout shell parameters as a function of the progenitor properties}
The analytic model that describes the emission as a function of the breakout shell properties, fits the results from analytic progenitors and from numerical progenitors, as described in sub-section \ref{subsection:numeric_lc}. In this sub-section we relate the breakout parameters to the three global parameters $M_{\rm ej}$, $R_*$ and $E_{\rm exp}$ for our set of progenitors. This will enable us later to find the light curve dependence on these parameters.

The definition of the breakout time that we use in order to determine the breakout parameters is the same as described in section \ref{section:semi_analytic}. At the breakout, $\tau_0=1.2c/v_0$ was obtained for the numerical stars, similarly to the analytic progenitors. The scaling of the breakout parameters for the numerical progenitors is similar to the analytic progenitors as well, except for the numerical coefficients and the $R_*$ dependency:

\begin{subequations}
\label{eq:numeric_hydro}
\begin{equation}
\label{eq:numeric_rho}
\rho_0 \approx 1.5 \cdot 10^{-9} \ \mathrm{g/cm^3} \ M_{15}^{0.67}R_{500}^{-0.64}E_{51}^{-0.31},
\end{equation}
\begin{equation}
\label{eq:numeric_v}
v_0 \approx 4500 \ \mathrm{km/s} \ M_{15}^{-0.44}R_{500}^{-0.49}E_{51}^{0.56},
\end{equation}
\begin{equation}
\label{eq:numeric_d}
\frac{d_0}{R_*} \approx 10^{-2} M_{15}^{-0.21}R_{500}^{-0.1}E_{51}^{-0.25}.
\end{equation}
\end{subequations}

\begin{figure}
    \centering
    \includegraphics[width=\linewidth]{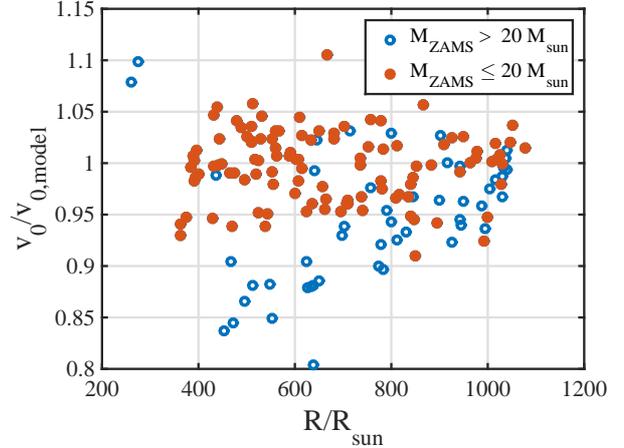}
    \caption{The values of $v_0$, obtained numerically for all the numerical progenitors, normalized to the scaling of equation \ref{eq:numeric_v}. Progenitors with $M_{\rm ZAMS} \leq 20$, which are more common, are marked with filled red circles, and the rest of the progenitors are marked with empty blue circles.}
    \label{fig:numeric_u_calibrated2}
\end{figure}

Since $R_*$ and $M_{\rm ej}$ are uncorrelated (see figure \ref{fig:numeric_progenitors}), the different $R_*$ dependence is explicit. In figure \ref{fig:numeric_u_calibrated2} the ratio between the numerical values of $v_0$ and the prediction of equation \ref{eq:numeric_v} is depicted for all the progenitors, as a function of the progenitor radius. As can be seen, for all progenitors with $M_{\rm ZAMS} \le 20M_\odot$ the scaling is good to within $10\%$ while for larger $M_{\rm ZAMS}$ progenitors it is accurate to within $20\%$.

\begin{figure}
    \centering
    \includegraphics[width=\linewidth]{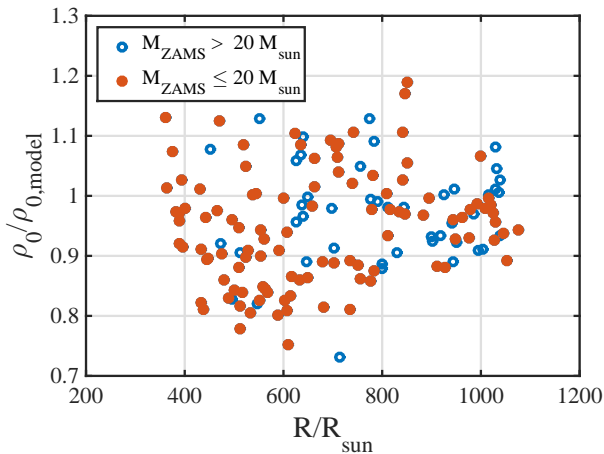}
    \caption{The values of $\rho_0$, obtained numerically for all the numerical progenitors, normalized to the scaling of equation \ref{eq:numeric_rho}. Symbols are the same as in figure \ref{fig:numeric_u_calibrated2}.}
    \label{fig:numeric_rho_calibrated2}
\end{figure}

\begin{figure}
    \centering
    \includegraphics[width=\linewidth]{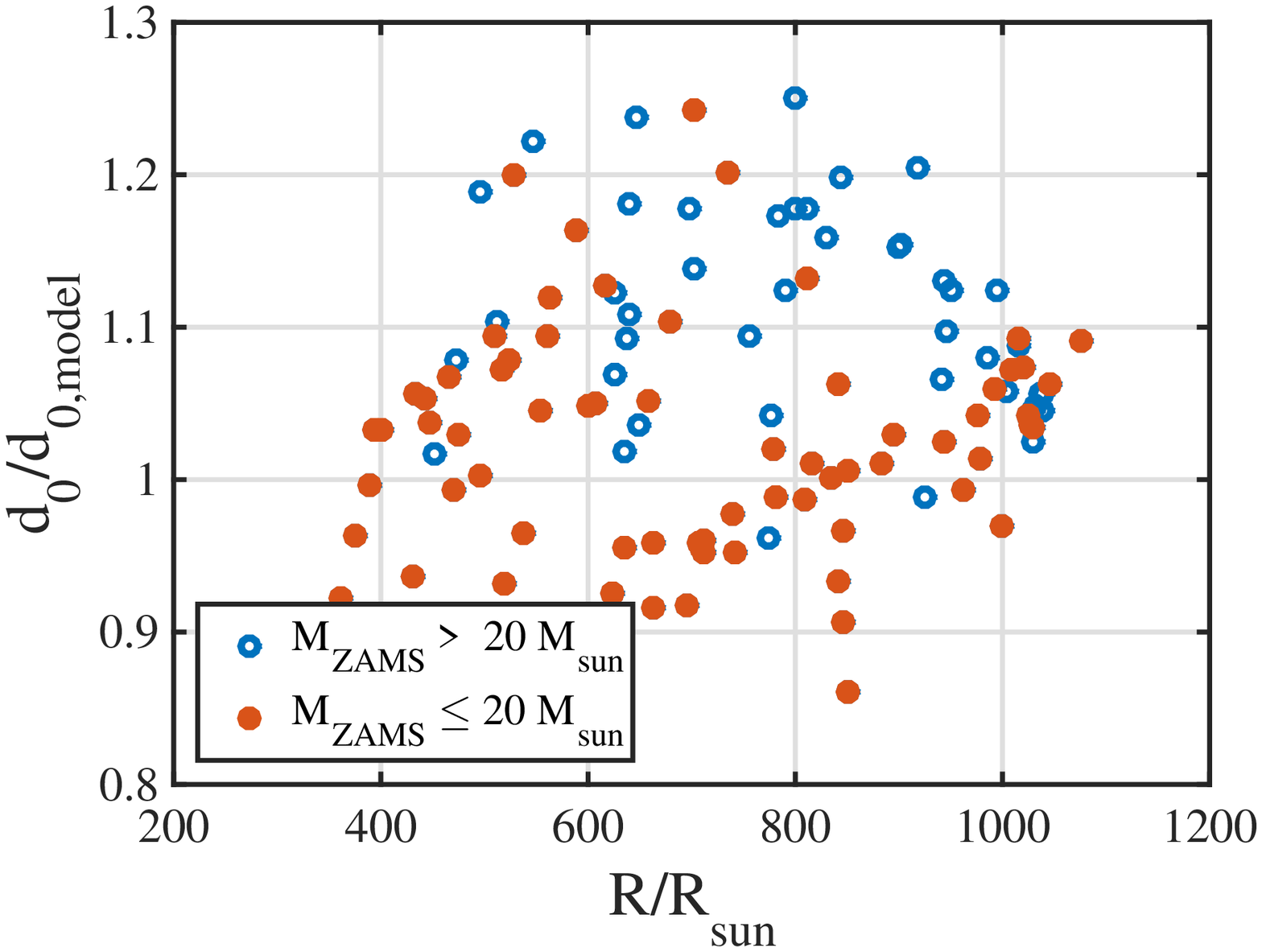}
    \caption{The values of $d_0$, obtained numerically for all the numerical progenitors, normalized to the scaling of equation \ref{eq:numeric_d}. Symbols are the same as in figure \ref{fig:numeric_u_calibrated2}.}
    \label{fig:numeric_d_calibrated2}
\end{figure}

Similar ratios for the breakout shell density and width are shown in figure \ref{fig:numeric_rho_calibrated2} and figure \ref{fig:numeric_d_calibrated2} respectively. The scaling is good to within $20\%$. The scaling of all the breakout properties with the explosion energy was checked using simulations of the same progenitor in different energies, and was found to be within $1\%$ (the expected numerical error).

\begin{figure}
    \centering
    \includegraphics[width=\linewidth]{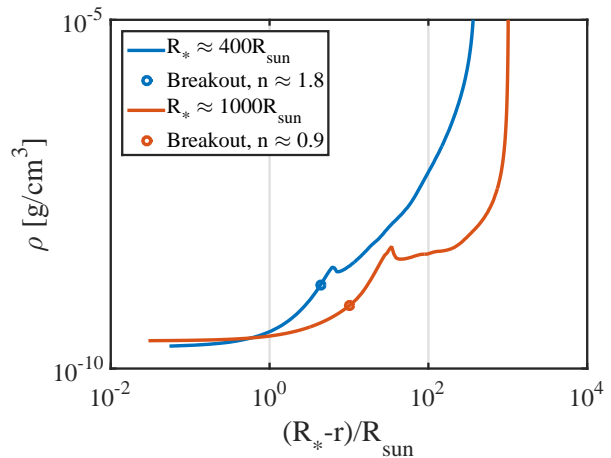}
    \caption{The density profiles of two numerical progenitors with a similar mass $M_{\rm ej}=12.5M_\odot$, as a function of the distance from the edge. The larger progenitor ($R_* \approx 1000R_\odot$) is characterized by $n'=0.9$ and the smaller one ($R_* \approx 400R_\odot$) is characterized by $n'=1.8$. In addition, the logarithmic index around the breakout shell varies less for the smaller progenitor, so it is better approximated as a power law density profile.}
    \label{fig:numeric_R_n}
\end{figure}

The difference between the analytic model, specified in equations \ref{eq:semi_analytic_hydro} and the numerical model of equations \ref{eq:numeric_hydro}, mainly in $R_*$ dependency, is due to the different stellar structure of progenitors with different radii. None of the progenitor's density profiles is exactly of a power law form, but, defining the density logarithmic index of the breakout shell, 
\begin{equation}
n' \equiv \frac{dlog(\rho)}{dlog(R_*-r)}\big|_{m=m_0},
\end{equation}
we find out that smaller progenitors tend to have higher values of $n'$. The value of $n'$ changes from around $1.8$ for progenitors with $R_*=400R_\odot$ to around $0.8$ for progenitors with $R_*=900R_\odot$. In addition, the logarithmic index around the breakout shell generally varies less for smaller progenitors. An example for these phenomena is given in figure \ref{fig:numeric_R_n}, in which two profiles of different progenitors, with similar masses and different radii are shown. This correlation between the progenitor radius and the mean and variance values of $n$ affects the scaling of the breakout properties with the progenitor radius.

Using equations \ref{eq:numeric_hydro}, it can be seen that equation \ref{eq:semi_analytic_tau} is not applicable for the numerical progenitors. Although $\tau_0=1.2c/v_0$, the relation $\tau \sim \kappa_T\rho_0d_0$ is correct only to a ($R_*$ dependent) factor of order of unity since $n'$ and the structure vary. The equivalent of equation \ref{eq:semi_analytic_tau} for our numerical progenitors set is
\begin{equation}
\label{eq:numeric_tau}
d_0 = \frac{2.75c}{\kappa_T\rho_0v_0R_{500}^{0.23}}.
\end{equation}

The dependence of the breakout shell temperature on $v_0$ and $\rho_0$ is the same as in equation \ref{eq:semi_analytic_Tbb} (within $3\%$ accuracy). Therefore, it is given by

\begin{equation}
\label{eq:numeric_Tbb}
T_{\rm{BB},0} \approx 5.4 \cdot 10^5 \ \mathrm{^\circ K} \ M_{15}^{-0.05}R_{500}^{-0.4}E_{51}^{0.2}.
\end{equation}

The scatter of the breakout shell temperature with respect to the model is determined by the scatter of $\rho_0^{0.25}v_0^{0.5}$, and is less than $10\%$.

\subsection{Condition for breakout in thermal equilibrium}
\label{subsection:thermal}

In our model for the observed temperature we assume that radiation is in thermal equilibrium. This is true at the time of the breakout when $\eta_0>1$, meaning that enough photons are generated within the breakout shell to maintain thermal equilibrium. If $\eta_0<1$, the emission is out of thermal equilibrium until at least the spherical phase (for further discussion see NS10). We can use our scaling of $\eta_0$ to provide an upper limit for the explosion energy for which the breakout is thermal. We substitute the scaling of the breakout properties from equation \ref{eq:numeric_hydro} into equation \ref{eq:semi_analytic_eta} and obtain

\begin{equation}
\label{eq:numeric_eta}
\eta_0 
%\approx 0.8\left(\frac{v_0}{10^4\rm{km/s}}\right)^\frac{15}{4}\left(\frac{\rho_0}{10^{-9}\rm{g/cm^3}}\right)^{-\frac{1}{8}}
= 4 \cdot 10^{-2} M_{15}^{-1.73}R_{500}^{-1.75}E_{51}^{2.14}.
\end{equation}
The maximal explosion energy for which the emission is in thermal equilibrium is therefore
\begin{equation}
\label{eq:numeric_maxinal_energy}
E_{51} = 4.5 M_{15}^{0.81}R_{500}^{0.82}.
\end{equation}
As can be seen, for a typical RSG explosion, the emission is indeed in thermal equilibrium from the breakout on.

\section{A calibrated analytic model of the bolometric luminosity and observed temperature}
\label{section:final}

In this section, we summarize the results of the previous sections and present a calibrated analytic model for the evolution of the luminosity and observed temperature. First (sub-section \ref{subsection:final_bo}) we provide a model that is based on the properties of the breakout shell, $v_0$, $\rho_0$ and $R_*$. This model is accurate and general in the sense that it is only weakly dependent on the exact properties of the progenitor structure. Then (\ref{subsection:final_prog}) we provide a model of the light curve as a function of the global explosion properties, $E_{\rm exp}$, $M_{\rm ej}$ and $R_*$. This model is slightly less accurate than the first one and it is also less general, as it depends on the correlations between properties in our set of numerical progenitors. Times are denoted  $t_{\mathrm{hr}}$, and $t_\mathrm{day}$ for units of seconds, hours and days respectively. The time of the bolometric emission peak when light travel time is not considered is defined as $t=0$. Bolometric and monochromatic light curves, calculated using the calibrated model presented in this section can be downloaded at \url{http://www.astro.tau.ac.il/~tomersh/}. 

\subsection{Bolometric light curve and observed temperature as functions of the breakout properties}
\label{subsection:final_bo}

The three parameters which characterise the breakout shell are $v_0$, $\rho_0$ and $R_*$, since $d_0$ is determined via equation \ref{eq:numeric_tau}. We denote $v_{0,x}=v_{0}/(x \cdot 10^3$ km/s), $\rho_{0,x}=\rho_{0}/(10^{x}$ g/cc) and $R_{x}=R_{*}/xR_\odot$.

The luminosity after the peak, neglecting light travel time, is obtained by substituting equations \ref{eq:semi_analytic_lc}, \ref{eq:numeric_tau} into equation \ref{eq:semi_analytic_L}:

\begin{multline}
\label{eq:final_L_bo}
\frac{L_{\rm obs}(t)}{1\mathrm{erg/s}} \simeq \\
\begin{cases}
1.6 \cdot 10^{45} v_{0,5}^{3} \rho_{0,-9}^{1} R_{500}^{2} & t \ll t_0 \\
3.2 \cdot 10^{43} v_{0,5}^{0.33} \rho_{0,-9}^{-0.33} R_{500}^{1.69} t_{\mathrm{hr}}^{-4/3} & t_0 \ll t \ll t_{\rm{s}} \\
3.3 \cdot 10^{42} v_{0,5}^{1.31} \rho_{0,-9}^{-0.33} R_{500}^{0.71} t_{\mathrm{day}}^{-0.35} & t_{\rm{s}} \ll t < t_{\rm rec}
\end{cases}.
\end{multline}
where $t_{\rm rec}$ is the time in which the observed temperature reaches $T=7500^\circ $K. At this point, recombination is no longer negligible, and the model described here is no longer valid. To obtain a smooth broken power law, the luminosity at the transitions between the phases is given by a sum, as described in equation \ref{eq:semi_analytic_L_transition}. During the breakout pulse ($t<0$), the luminosity is obtained by substituting equation \ref{eq:semi_analytic_lc} into equation \ref{eq:semi_analytic_rise}:

\begin{multline}
\label{eq:final_L_rise_bo}
\frac{L_{\rm obs}(t)}{1\mathrm{erg/s}} \simeq \\
1.6 \cdot 10^{45} v_{0,5}^{3} \rho_{0,-9}^{1} R_{500}^{2}e^{-0.35(t/t_0)^2+0.15(t/t_0)}.
\end{multline}

The observed temperature after the peak is obtained by substituting equations \ref{eq:semi_analytic_lc}, \ref{eq:semi_analytic_eta} and \ref{eq:semi_analytic_tc} into equation \ref{eq:semi_analytic_T}: 

\begin{multline}
\label{eq:final_T_bo}
\frac{T_{\rm obs}(t)}{1\mathrm{^\circ K}} \simeq \\
\begin{cases}
4.2 \cdot 10^{5} v_{0,5}^{0.76} \rho_{0,-9}^{0.24} & t < t_0 \\
1.1 \cdot 10^{5} v_{0,5}^{-0.13} \rho_{0,-9}^{-0.21} R_{500}^{-0.1} t_{\mathrm{hr}}^{-0.45} & t_0 \leq t < t_{\rm{s}} \\
3.3 \cdot 10^{4} v_{0,5}^{-0.03} \rho_{0,-9}^{-0.2} R_{500}^{-0.2} t_{\mathrm{day}}^{-0.35} & t_{\rm{s}} \leq t < t_{\rm{c}} \\
4.1 \cdot 10^{4} v_{0,5}^{-0.55} \rho_{0,-9}^{-0.18} R_{500}^{0.06} t_{\mathrm{day}}^{-0.6} & t_{\rm{c}} \leq t < t_{\rm rec}
\end{cases}.
\end{multline}

Before the peak, the observed temperature is roughly constant. The observed spectrum is not blackbody, since higher energy photons are emitted from inner shells with higher temperatures. The observed spectrum is well described by equations \ref{eq:T_col}, \ref{eq:spectrum_nobb} until the peak of the spectrum, where line blanketing becomes significant. The transition times, and recombination time are given by
\begin{subequations}
\label{eq:final_prop_bo}
\begin{equation}
\label{eq:final_t0_bo}
t_0 = 190 \ \mathrm{s} \ v_{0,5}^{-2} \rho_{0,-9}^{-1} R_{500}^{-0.23},
\end{equation}
\begin{equation}
t_{\rm{s}} = 3.2 \ \mathrm{hr} \ v_{0,5}^{-1} R_{500}^{1},
\end{equation}
\begin{equation}
t_{\rm{c}} = 2.5 \ \mathrm{day} \ v_{0,5}^{-2.07} \rho_{0,-9}^{0.08} R_{500}^{1.06},
\end{equation}
\begin{equation}
\label{eq:final_trec_bo}
t_{\rm rec} = 17 \ \mathrm{day} \ v_{0,5}^{-0.92} \rho_{0,-9}^{-0.31} R_{500}^{0.1}.
\end{equation}
\end{subequations}

This model describes the luminosity for all numerically calculated progenitors to within $15\%$ ($30\%$ late in the spherical phase) and the observed temperature to within $10\%$. This inaccuracy is due to the different structure of the progenitors.

In order to include light travel time, one should convolve the results given in this section with equation \ref{eq:ltt_general}. The peak luminosity including light travel time is obtained by substituting equations \ref{eq:numeric_tau} and \ref{eq:semi_analytic_lc}  into equation \ref{eq:numeric_ltt_L}:
\begin{equation}\label{eq:L0ltt_BOparam}
L_{0,ltt} = 8 \cdot 10^{44} \ \mathrm{erg} \ v_{0,5} R_{500}^{0.77} f.
\end{equation}
$f$ is a numerical factor of order $1$, given by equation \ref{eq:numeric_ltt_f}. With light travel time included, the duration of the rise is of order $t_{\rm Rc}=R_*/c$, while the initial rapid exponential rise is still characterised by a time scale $t_0$. In addition, around the peak, the spectrum is not thermal, but characterised by a range of temperatures. For further discussion, see sub-section \ref{subsection:ltt}. A useful ratio to assess the importance of light travel time is
\begin{equation}
t_0/t_{\rm Rc} = 0.16 v_{0,5}^{-2} \rho_{0,-9}^{-1} R_{500}^{-1.23}.
\end{equation}

Examination of equations \ref{eq:final_L_bo}-\ref{eq:L0ltt_BOparam} shows that many of the observables (i.e., characteristic time scales and the luminosity and temperature at different regimes) depend strongly on $v_0$ and $R_*$. Therefore early observations will tightly constrain both parameters. However, only the rise time of the breakout pulse $t_0$, depends strongly on $\rho_0$. Without observations of the breakout with a temporal resolution of a minute or so, it will be hard to tightly constrain $\rho_0$.  

\subsection{Bolometric light curve and observed temperature as functions of global SN properties}
\label{subsection:final_prog}

The three parameters which we use to characterise the SN are $R_*$, $M_{\rm ej}$ and $E_{\rm exp}$. For the numerically calculated progenitors, the dependency of the breakout properties on these parameters is specified in equation \ref{eq:numeric_hydro}. We denote $M_{x}=M_{\rm ej}/xM_\odot$, $R_{x}=R_*/xM_\odot$ and $E_{x}=E_{\rm exp}/10^x$erg.

The luminosity after the peak, neglecting light travel time, is obtained by substituting equations \ref{eq:numeric_hydro} into equation \ref{eq:final_L_bo}:

\begin{multline}
\label{eq:final_L}
\frac{L_{\rm obs}(t)}{1\mathrm{erg/s}} \simeq \\
\begin{cases}
1.8 \cdot 10^{45} M_{15}^{-0.65}R_{500}^{-0.11}E_{51}^{1.37} & t \ll t_0 \\
2.7 \cdot 10^{43} M_{15}^{-0.34}R_{500}^{1.74}E_{51}^{0.29} t_{\mathrm{hr}}^{-4/3} & t_0 \ll t \ll t_{\rm{s}} \\
1.6 \cdot 10^{42} M_{15}^{-0.78}R_{500}^{0.28}E_{51}^{0.84} t_{\mathrm{day}}^{-0.35} & t_{\rm{s}} \ll t < t_{\rm rec}
\end{cases}.
\end{multline}

Again, at the transitions between the phases, the luminosity is given by equation \ref{eq:semi_analytic_L_transition}. Similarly, the luminosity during the breakout pulse ($t<0$) obeys
\begin{multline}
\label{eq:final_L_rise}
\frac{L_{\rm obs}(t)}{1\mathrm{erg/s}} \simeq \\
1.8 \cdot 10^{45} M_{15}^{-0.65}R_{500}^{-0.11}E_{51}^{1.37}e^{-0.35(t/t_0)^2+0.15(t/t_0)}.
\end{multline}

The observed temperature after the peak is
\begin{multline}
\label{eq:final_T}
\frac{T_{\rm obs}(t)}{1\mathrm{^\circ K}} \simeq \\
\begin{cases}
4.3 \cdot 10^{5} M_{15}^{-0.17}R_{500}^{-0.52}E_{51}^{0.35} & t < t_0 \\
1 \cdot 10^{5} M_{15}^{-0.07}R_{500}^{0.1}E_{51}^{-0.01} t_{\mathrm{hr}}^{-0.45} & t_0 \leq t < t_{\rm{s}} \\
3 \cdot 10^{4} M_{15}^{-0.11}R_{500}^{-0.04}E_{51}^{0.04} t_{\mathrm{day}}^{-0.35} & t_{\rm{s}} \leq t < t_{\rm{c}} \\
4.1 \cdot 10^{4} M_{15}^{0.13}R_{500}^{0.46}E_{51}^{-0.25} t_{\mathrm{day}}^{-0.6} & t_{\rm{c}} \leq t < t_{\rm rec}
\end{cases}.
\end{multline}

Before the peak, the observed temperature is roughly constant. The observed spectrum is well described by equations \ref{eq:T_col}, \ref{eq:spectrum_nobb} up to frequencies higher than the peak of the spectrum, where line blanketing becomes significant. The transition times, and recombination time are given by
\begin{subequations}
\label{eq:final_prop}
\begin{equation}
t_0 = 155 \ \mathrm{s} \ M_{15}^{0.23}R_{500}^{1.39}E_{51}^{-0.81},
\end{equation}
\begin{equation}
t_{\rm{s}} = 3.6 \ \mathrm{hr} \ M_{15}^{0.44}R_{500}^{1.49}E_{51}^{-0.56},
\end{equation}
\begin{equation}
t_{\rm{c}} = 3.2 \ \mathrm{day} \ M_{15}^{0.97}R_{500}^{2.02}E_{51}^{-1.19},
\end{equation}
\begin{equation}
\label{eq:final_trec}
t_{\rm rec} = 16.6 \ \mathrm{day} \ M_{15}^{0.22}R_{500}^{0.76}E_{51}^{-0.43}.
\end{equation}
\end{subequations}

This model describes the luminosity for all numerical progenitors to within $25\%$ during the first day and $35\%$ at later times. The observed temperature is described to within  $15\%$. The inaccuracy is larger than the inaccuracy of the breakout properties model, since the scaling of the breakout properties with the progenitor properties is only accurate to within $20\%$. This model is also less general since it depends on the specific structures of the progenitors we calculated.  %The scaling with the explosion energy is accurate to within $1\%$ (the numerical error).

The peak luminosity including light travel time is
\begin{equation}\label{eq:L0ltt_EMR}
L_{0,ltt} = 7.2 \cdot 10^{44} \ \mathrm{erg} \ M_{15}^{-0.42}R_{500}^{0.28}E_{51}^{0.56} f,
\end{equation}
where $f$ is a numerical factor of order $1$, given by equation \ref{eq:numeric_ltt_f} (see discussion at section \ref{subsection:ltt}). The ratio between the diffusion time at breakout and the light travel time is
\begin{equation}
t_0/t_{\rm Rc} = 0.15 M_{15}^{0.23}R_{500}^{0.39}E_{51}^{-0.81}.
\end{equation}

In the previous subsection (\ref{subsection:final_bo}) we deduced that all the observables, except for $t_0$, depend mostly on two parameters ($v_0$ and $R_*$). This is also the case here where the observables depend strongly on $R_*$ and $E_{\rm exp}/M_{\rm ej}$ (which is tightly related to $v_0$) but weakly on $E_{\rm exp}$ and $M_{\rm ej}$ separately. The only exception, again, is $t_0$. Therefore, detailed early light curve observations will provide tight constraints on $R_*$ and $E_{\rm exp}/M_{\rm ej}$, but unless the rise of the breakout pulse, which is on a time scale of minutes, is resolved, it will be hard to constrain $E_{\rm exp}$ and $M_{\rm ej}$ separately.  

\section{Properties of the optical \& UV light curve}
\label{section:optical}

\begin{figure}
    \centering
    \includegraphics[width=\linewidth]{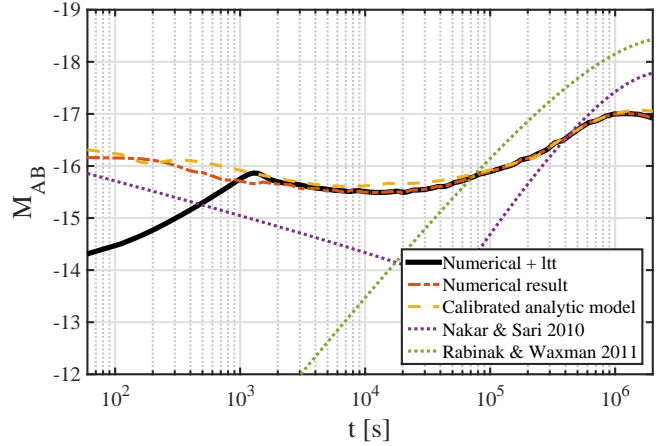}
    \caption{The R-band ($\lambda=640$nm) light curve, for the same numerical progenitor as used in figure \ref{fig:numeric_lc_calibrated}. The numerical result with (without) light travel time is plotted in a solid black (dotted-dashed red) line. It is compared to the analytic model (dashed orange line), which is obtained by substituting equations \ref{eq:final_L}, \ref{eq:final_T} into equation \ref{eq:spectrum_nobb}. Also plotted are the models of NS10 (dotted purple line) and RW11 (dotted green line).}
    \label{fig:optic2_lc_R}
\end{figure}

\begin{figure}
    \centering
    \includegraphics[width=\linewidth]{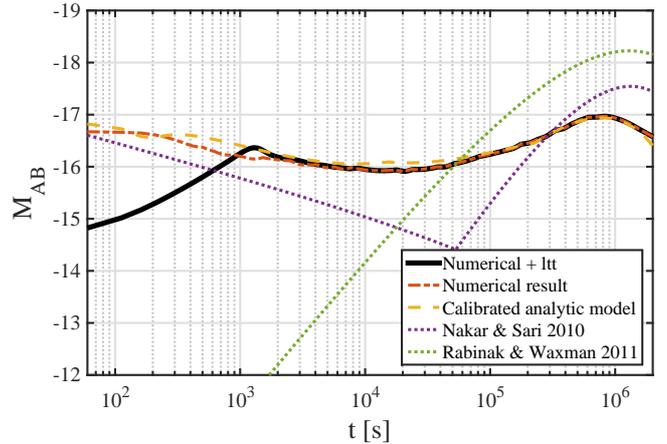}
    \caption{The {\it g}-band ($\lambda=470$nm) light curve, for the same numerical progenitor as used in figure \ref{fig:numeric_lc_calibrated}. Symbols are the same as in figure \ref{fig:optic2_lc_R}.}
    \label{fig:optic2_lc_G}
\end{figure}

In this section we discuss the properties of the optical \& UV light curve at early times. A monochromatic light curve can be obtained for every wavelength, by substituting $L,T$ from equations \ref{eq:final_L} - \ref{eq:final_T} into equation \ref{eq:spectrum_nobb}. Equation \ref{eq:spectrum_nobb} is inaccurate for photon energies above those of the blackbody peak ($h\nu \gtrsim 3k_BT$) due to line blanketing. Therefore, the monochromatic light curves presented here are inaccurate at times after the peak is seen in the specific observed wavelength, as line blanketing causes the decline after the peak to be faster than predicted by our model.

\begin{figure}
    \centering
    \includegraphics[width=\linewidth]{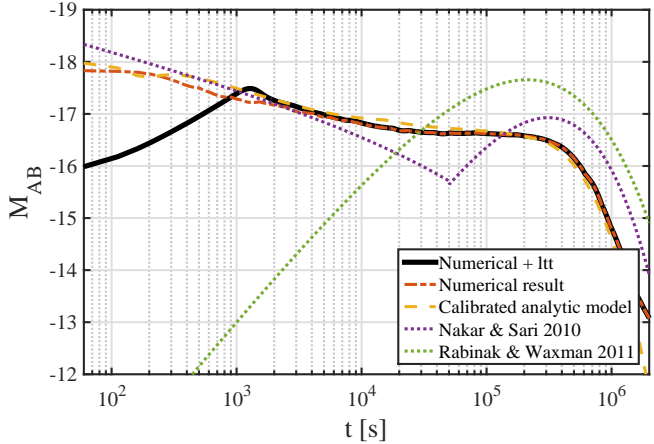}
    \caption{The near-UV ($\lambda=200$nm) light curve, for the same numerical progenitor as used in figure \ref{fig:numeric_lc_calibrated}. Symbols are the same as in figure \ref{fig:optic2_lc_R}.}
    \label{fig:optic2_lc_UV}
\end{figure}

\subsection{Emission after the breakout pulse}

We give examples of the emission in R-, {\it g}-, and near-UV bands. Numerical light curves for R-band ($\lambda=640$nm), with and without light travel time, are plotted in figure \ref{fig:optic2_lc_R}. The calculation was performed with the same progenitor as in figure \ref{fig:numeric_lc_calibrated}, but the observed features are similar for all the progenitors in our sample. This result is compared to the optical emission obtained by substituting the model for $L$ and $T$ (section \ref{section:final}) into equation \ref{eq:spectrum_nobb}, and to the models of NS10 and RW11.

Our model fits the calculation (neglecting light travel time) within $0.2$ mag. The emission is double peaked, as expected in NS10, although the first peak is less pronounced compared to their prediction. During the planar phase, the optical bands are in the modified Rayleigh-Jeans regime where\footnote{NS10 predicted $T \propto t^{-0.35}$ during the planar phase, while we find $T \propto t^{-0.45}$. However, NS10 assumed a blackbody spectrum which yields $L_\nu \propto L/T^{3} \propto t^{-0.3}$ during the planar phase, so the decline is similar.} $L_\nu \propto L/T^{2.4} \propto t^{-0.25}$. By the beginning of the spherical phase, the luminosity falls more slowly and the temperature falls faster, thus the monochromatic luminosity rises ($L/T^{2.4} \propto t^{0.5}$). After $t=t_{\rm{c}}$ the rise is even more rapid ($L/T^{2.4} \propto t^{1}$), though by this time most of the wavelengths are not purely in the Rayleigh-Jeans regime. Light travel time affects the emission at times earlier than $t_{\rm Rc}-t_{0}$, which is the peak time of the emission with light travel time included.

Similar light curves for the same progenitor are depicted in figure \ref{fig:optic2_lc_G} at {\it g}-band ($\lambda=470$nm) and in figure \ref{fig:optic2_lc_UV} at near-UV ($\lambda=200$nm). The {\it g}-band light curves show features similar to that of the R-band. The main difference is that the rise is less rapid and the peak is seen a few days earlier (see discussion below). For the near-UV, only one peak exists, followed by a slow decline that becomes after $T_{\rm obs}$ falls below the observed band a very fast decline. This feature is generic for most of the numerical progenitors. Since our model does not account for line blanketing, we expect a more rapid decrease than predicted by our blackbody model after the end of the slow decrease ($t \approx 5$ days) in the near-UV emission.

\begin{figure}
    \centering
    \includegraphics[width=\linewidth]{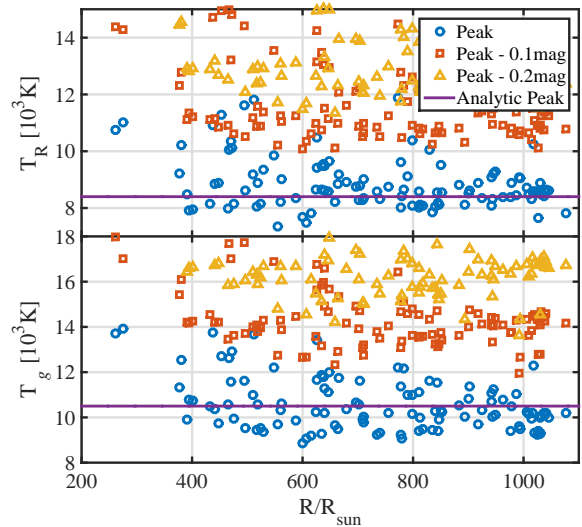}
    \caption{Top: the observed temperature at the peak of the R-band ($\lambda=640$nm) as a function of the stellar radius (blue circles), calculated numerically for all the numerical progenitors. The temperature when the luminosity is lower by $0.1$ mag (red rectangles) and by $0.2$ mag (yellow triangles) is also shown. The purple line depicts the analytic prediction $x_{\rm peak}=2.7$. Bottom: the same for the {\it g}-band ($\lambda=470$nm), where $x_{\rm peak}=2.9$}
    \label{fig:optic_T_vs_R}
\end{figure}

The second peak in the optical bands is seen when the observed band approaches the peak of the spectrum due to the decrease in the observed temperature. This happens at earlier times for bluer bands. By this time, the observed band is no longer in the Rayleigh-Jeans regime. The value of $x \equiv h\nu / k_BT_{\rm obs}$ at the peak can be estimated based on equation \ref{eq:spectrum_nobb} and depends on the evolution of $T$ and $L$. For the late spherical phase $t \geq t_{\rm{c}}$ our model predicts $x_{\rm peak}=2.95$. For both R- and {\it g}-band light curves, the peak is reached at times later than $t_{\rm{c}}$. The peak time is obtained by substituting equation \ref{eq:final_T} into $T(t)=h\nu/k_Bx_{\rm peak}$:
\begin{equation}
\label{eq:t_peak}
\frac{t_{\rm peak}}{1\mathrm{day}} = 4.9 M_{15}^{0.22}R_{500}^{0.76}E_{51}^{-0.43}\nu_{15}^{-1.67}\left(\frac{x_{\rm peak}}{2.95}\right)^{1.67}.
\end{equation}

\begin{figure}
    \centering
    \includegraphics[width=\linewidth]{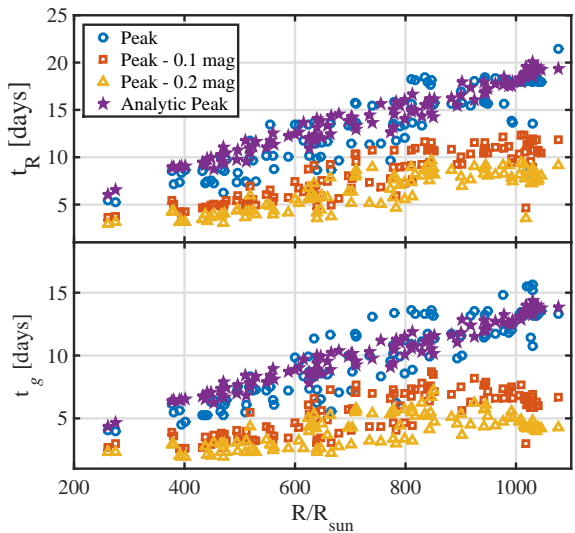}
    \caption{Top: the peak time (blue circles), and the times where the luminosity is lower by $0.1$ mag (red squares) and $0.2$ mag (yellow triangles), for the R-band ($\lambda=640$nm) light curve as a function of the stellar radius. The times are calculated numerically for all the numerical progenitors. The analytic prediction (see equation \ref{eq:t_peak}) is marked by purple pentagrams. Bottom: the same for the {\it g}-band ($\lambda=470$nm)}
    \label{fig:optic_time_vs_R}
\end{figure}

The $R_*$ dependence of $t_{\rm peak}$ is similar to the result of \citet{Morozova2016} (equation 4) which found $t_{\rm peak} \propto R_*^{0.82}$. Figure \ref{fig:optic_T_vs_R} depicts the temperature at the peak of R- (top) and {\it g}- (bottom) bands, and at $0.1$ mag and $0.2$ mag fainter than the peak, for all the progenitors. For the {\it g}-band, the peak temperature is around $T=10000^\circ $K, which corresponds to $x_{\rm peak}=2.9$, while for the R-band the peak temperature corresponds to $x_{\rm peak}=2.7$. The slightly lower values of $x_{\rm peak}$, compared to the model, are due to the steeper luminosity decrease near the time of recombination (figure \ref{fig:numeric_all_lc}). Note that the temperature is significantly higher than at the peak, by about $25\%$ [$40\%$], when the optical luminosity is only $0.1$ [$0.2$] mag fainter. In figure \ref{fig:optic_time_vs_R} the peak time and the times where the luminosity is lower by $0.1$ mag and $0.2$ mag are shown. The ratio between the numerical peak time and the analytic model prediction (equation \ref{eq:t_peak}) is between $0.8$ and $1.1$.

Our results highlight the difficulty in constraining the progenitor properties from current observations at a single band. The resolution of observations makes it hard to determine accurately the time of the exact peak, and even an error of $0.1-0.2$ mag in the peak magnitude can lead to an error of $50\%$ in the estimation of $R_*$. However, since the temperature evolves significantly before the peak, and its value at the peak depends only on the observed band, spectral or multi-wavelength observations can be used to significantly improve the constraints.

\subsection{The breakout pulse}

\begin{figure}
    \centering
    \includegraphics[width=\linewidth]{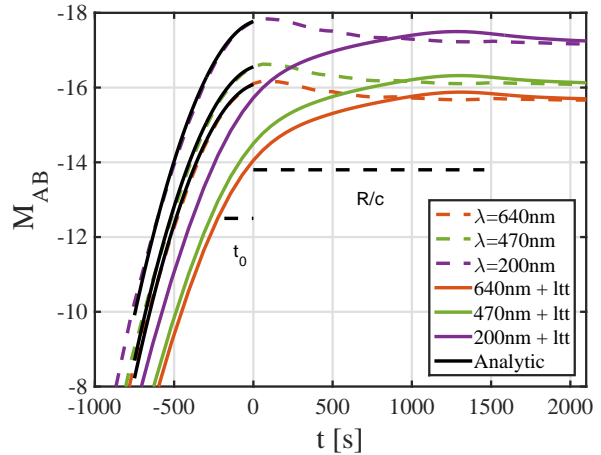}
    \caption{R-band (red line), {\it g}-band (green line) and near-UV (purple line) emission at early times, for the same numerical progenitor as used in figure \ref{fig:numeric_lc_calibrated}. The numerical result with (without) light travel time is plotted in solid (dashed) lines. Including light travel time, the very early rise is of a typical time $t_0$ while the rise near the peak is of a typical time $t_{\rm Rc}$. The analytic model (equation \ref{eq:final_L_rise}) (solid black line) is depicted in all wavelengths.}
    \label{fig:optic2_lc_rise}
\end{figure}

As discussed in sub-section \ref{subsection:ltt}, light travel time affects the emission during the breakout pulse. Figure \ref{fig:optic2_lc_rise} shows the rise of the R-, {\it g}- and near-UV bands. The analytic model (equation \ref{eq:final_L_rise}) fits the rise very well (light travel time neglected). The emission including light travel time is composed of a rapid increase with a typical time $t_0$ and a slower increase with a typical time $t_{\rm Rc}$.

\section{The photospheric velocity}
\label{section:photosphere}

Additional information can be extracted from the velocity of observed lines in early spectra. In particular, specific lines are considered as good estimators of the photospheric velocity. These are harder to identify in early spectra compared to late ones. Nevertheless, we provide here an analytic model for the photospheric velocity at early times (and a comparison to the numerical results), for cases where it can be estimated before recombination becomes significant.

The position of the photosphere is determined by the point which satisfies $\tau=1$. Since the optical depth and the shell mass, both measured from the edge of the star, are related via $\tau \approx \kappa m / 4\pi R^2$, during the planar phase the photosphere is located at $m \approx 4 \pi R_*^2 / \kappa$. During the spherical phase, however, $R(m) \approx 2v(m)t$ \citep[the factor of $2$ is due to the rarefaction, ][]{Matzner1999}. The analytic prediction is $v(m) \propto m^{-0.12}$ (for $n=1.5$ with weak dependency on $n$), which in turn yields $m_{ph} \propto t^{1.63}$. The best fit of the numerical results to a piecewise power law is similar to this analytic prediction, and is
\begin{multline}
\label{eq:m_phot}
\frac{m_{ph}(t)}{M_\odot} \simeq \\
\begin{cases}
2.27 \cdot 10^{-5} R_{500}^{2} & t < 1.5t_{\rm{s}} \\
1.45 \cdot 10^{-6} M_{15}^{-0.71}R_{500}^{-0.43}E_{51}^{0.91} t_{\mathrm{hr}}^{1.63} & 1.5t_{\rm{s}} \leq t
\end{cases},
\end{multline}
and 
\begin{multline}
\label{eq:R_phot}
\frac{R_{ph}(t)}{R_\odot} \simeq \\
\begin{cases}
R_* & t < 1.5t_{\rm{s}} \\
128 M_{15}^{-0.36}R_{500}^{-0.21}E_{51}^{0.45} t_{\mathrm{hr}}^{0.81} & 1.5t_{\rm{s}} \leq t
\end{cases}.
\end{multline}

\begin{figure}
    \centering
    \includegraphics[width=\linewidth]{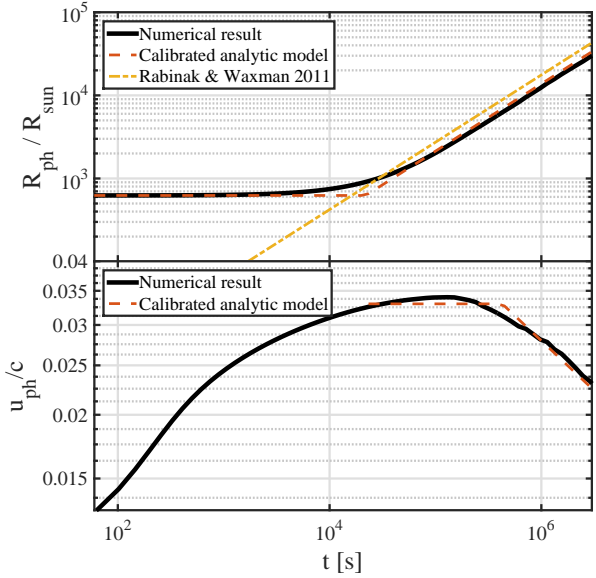}
    \caption{Top: the radius of the photosphere as a function of time from the peak luminosity, for the same numerical progenitor as used in figure \ref{fig:numeric_lc_calibrated}. The numerical result (solid black line) is compared to the analytic model (equation \ref{eq:R_phot}, dashed red line) and to the analytic model of RW11 (dotted-dashed green line). Bottom: The velocity at the photosphere as a function of time from the peak luminosity. The numerical result (solid black line) is compared to the analytic model (equation \ref{eq:u_phot}, solid red line).}
    \label{fig:numeric_photosphere}
\end{figure}

The spherical phase of the photosphere evolution begins at $1.5t_{\rm{s}}$ where $t_{\rm{s}}$ is given in equation \ref{eq:final_prop}. A typical evolution of $R_{ph}$, for the same numerical progenitor from figure \ref{fig:numeric_lc_calibrated}, is depicted at the top of figure \ref{fig:numeric_photosphere}. The analytic model of equation \ref{eq:R_phot} is compared to the numerical result, and is found to agree to within $5\%$ during the planar and spherical phases, and to within $25\%$ at the transition point. The results are also compared to the model of RW11 (see equation 12 therein), which yields a radius larger by about $35\%$ during the spherical phase, relative to the calculation.

The numerical photosphere velocity is shown at the bottom of figure \ref{fig:numeric_photosphere}. The photosphere accelerates during the whole planar phase. At the beginning of the spherical phase, while the photosphere is still in the breakout shell, the velocity profile $v(m) \approx 2v_0$ is about constant (and accelerates slowly with time). Therefore, $u_{ph}$ is almost constant. As soon as $m_{ph}(t) \geq m_{0}$, the velocity profile is well approximated by $v(m)=m^{-0.12}$ and $u_{ph} \propto t^{0.2}$. Again, the best fit to the numerical results is similar to the analytic prediction:

\begin{multline}
\label{eq:u_phot}
\frac{u_{ph}(t)}{1\mathrm{km/s}} \simeq \\
\begin{cases}
9 \cdot 10^{3} M_{15}^{-0.44}R_{500}^{-0.49}E_{51}^{0.56} & 1.5t_{\rm{s}} < t < t_m \\
1.2 \cdot 10^{4} M_{15}^{-0.3}R_{500}^{-0.14}E_{51}^{0.38} t_{\mathrm{day}}^{-0.2} & t_m \leq t
\end{cases}.
\end{multline}
Here, $t_m$ is the time for which $m_{ph}(t_m)=m_0$:
\begin{equation}
t_m = 4.5 \ \mathrm{day} \ M_{15}^{0.71}R_{500}^{1.79}E_{51}^{-0.9}.
\end{equation}

This model fits the position and velocity of the photosphere to within $15\%$ accuracy (except for the transition point around $t=t_m$) for all the numerical progenitors.

\section{Summary}
\label{section:discussion}

We have studied the emission from a SN generated by the core-collapse of a RSG during the first $10-20$ days after first light, when recombination is negligible, and the light curve is mainly determined by the thermal energy distribution at the outer envelope. We used a 1D hydro-radiation code to simulate early light curves from explosions of different RSG progenitors, and combined analytic estimates with the numerical results to obtain an accurate analytic model for the emission. We first studied the effect of each of the SN properties $R_*$, $M_{\rm ej}$ and $E_{\rm exp}$ independently by defining an analytic prototype progenitor profile, and varying each parameter separately over a wide range of values. Then, we simulated the explosions of $124$ RSG progenitors calculated using the stellar evolution code MESA. The numerically calculated  progenitors are more realistic and have different profiles but the effect of $R_*$ and $M_{\rm ej}$ cannot be studied independently over a large range of values.

First, we found that earlier analytic works deviate from the numerical results by a factor of $2-4$ describing the bolometric luminosity and the features of the light curve, and by up to $50\%$ describing the observed temperature. This deviation is mostly due to the approximations made regarding the hydrodynamic evolution of the explosion, the initial progenitor density profile and the opacity of the matter. Then, we constructed a new calibrated analytic model which describes the light curve as a function of the breakout shell properties ($v_0$, $\rho_0$ and $R_*$) and as a function of the progenitor properties ($M_{ej}$ and $R_*$) and explosion energy. Our model is analytic, hence the dependency on each parameter is clearly understood, yet it has the advantage of being accurate, since it was calibrated by numerical simulations where most of the assumptions and approximations made in previous analytic models are relaxed.

We found that the dependency of the light curve on the breakout shell properties is of a global nature. Progenitors with very different internal structures but the same breakout parameters produce almost identical light curves during the first day and show only minor deviation at later times. This is because during the first day (up to the spherical phase) only emission from the breakout shell is observed. Later, during the spherical phase, inner parts of the progenitor are observed but the dependence of the emission on the exact structure is mild. Thus, the early light curves directly probe mostly the properties of the breakout shell. Our model relates the bolometric luminosity to the breakout shell properties within an accuracy of $15\%$ during the first day and at an accuracy of $25\%$ at later times. The observed temperature is predicted by the analytic model within an accuracy of about $10\%$ at all times. The early light curve is especially sensitive to $v_0$ and $R_*$, therefore these are the two parameters that can be most easily extracted from an early observation. The value of $\rho_0$ strongly affects only the timescale of the rise of the breakout emission (typically of order of minutes), therefore its value can not be well constrained without a detailed observation of the rise.

Global properties such as the ejecta mass and the explosion energy are not probed directly, but through their relation to the breakout properties. Since the mapping between $E_{\rm exp}$, $M_{ej}$ and $R_*$ and the breakout parameters depends on the progenitor structure, the relations that we provide between the early light curve and  $E_{\rm exp}$, $M_{ej}$ and $R_*$ are less accurate than the relations with the breakout parameter. They are also less general since they depend on the specific set of numerically calculated progenitors that we explored. Our model relates the bolometric luminosity to the progenitor and explosion properties within an accuracy of $25\%$ during the first day and $35\%$ at later times. The observed temperature is described to within $15\%$ accuracy.

We also derived an analytic approximation for the deviation of the observed spectrum from blackbody, mainly at the Rayleigh-Jeans regime. The source of this deviation is the higher absorption opacity at lower frequencies. As a result, photons with different frequencies are generated in different locations within the outflow, as less energetic photons are generated at outer locations where the electrons are colder. We show that over a limited range of frequencies below the spectral peak ($h\nu < k_BT_{\rm obs}$), which for typical parameters includes the UV and optical bands, the spectrum can be approximated as $L_\nu \propto \nu^{1.4}$. This deviation has a significant effect on the optical/UV light curve during the first day, when $T_{\rm obs} \gg 10^4~ ^\circ K$, and a lesser effect at later times.

We used our results to derive and explore optical and near-UV light curves. Similarly to previous analytic (NS10) and numerical \citep{Tominaga2011} results, the optical light curves depict two peaks. The first one corresponds to the breakout pulse (time scale of $R_*/c$) and is less prominent than predicted by NS10. The second (time scales of days) is the one observed in many SNe \citep[e.g.,][]{Anderson2014,Gonzales2015,Gall2015,Rubin2015} and corresponds to the passage of the spectral peak through the observed band. The shape of the UV light curve, however, is different than previously predicted. Only the first peak (time scale of $R_*/c$) is observed, and followed by a very slow decline for several days, which turns into a very fast decline when the spectral peak drops below the observed UV band.

The time of the second optical peak was recently used to constrain the progenitor radius (\citealt{Gonzales2015}; see however \citealt{Rubin2015}). We found a relation between the time of this peak and the global SN parameters, and showed that it is most sensitive to the progenitor radius, in agreement with previous studies (e.g., NS10, RW11, \citealt{Tominaga2011}). However, we showed that deriving constraints from the time of the peak in a single band observation is very sensitive to the exact identification of its location. For example, identifying the peak at the time that the flux is only $0.1-0.2$ mag fainter than the actual peak can lead an error of about $50\%$ in the estimation of $R_*$. This highlights the difficulty in constraining progenitor properties from current observations at a single band. We also examined $T_{\rm obs}$ at the time of the peak and found that it is higher, but not by much, than the recombination temperature. $T_{\rm obs}$ vary rapidly near the peak, and it is higher than at the peak, by about 25\% [40\%], when the optical luminosity is only 0.1 [0.2] mag fainter than the peak. This shows that much better constraints on the SN properties can be obtained with an information about the temperature near the peak (e.g., via multi-band observations).

To conclude, we present an accurate analytic model for the early emission of type II SNe that can be used for analyzing large data sets, planning future observations, or constraining progenitor properties from a given observation. Light curves generated using our calibrated model can be downloaded at \url{http://www.astro.tau.ac.il/~tomersh/}.
%For the latter goal, the main features that can be observed today are $t_{\rm{c}}$, $t_{\rm peak}$ and $L_{\nu,\rm peak}$. Measuring these provides independent constraints on $R_*$ and $E_{\rm exp}/M_{\rm ej}$. Future observations with temporal resolution of hours can be used to also measure $t_{\rm{s}}$ which provides another constrain on $R_*$ or $E_{\rm exp}/M_{\rm ej}$. Only if a temporal resolution of minutes is reached, $t_0$ can be measured and a third independent constraint on $M_{\rm ej}^2/E_{\rm exp} R_*^2$ is obtained. Then, each of the progenitor parameters is constrained independently.

\section*{Acknowledgements}

We are grateful to Nir Sapir for stimulating discussions, and to Sivan Ginzburg for his helpful advice regarding the numerical simulation. TS and EN were partially supported by an ERC starting grant (GRB/SN), ISF grant (1277/13) and an ISA grant.

%%%%%%%%%%%%%%%%%%%%%%%%%%%%%%%%%%%%%%%%%%%%%%%%%%



\bibliographystyle{mnras}
\bibliography{refs} % if your bibtex file is called example.bib

%%%%%%%%%%%%%%%%%%%%%%%%%%%%%%%%%%%%%%%%%%%%%%%%%%

%%%%%%%%%%%%%%%%% APPENDICES %%%%%%%%%%%%%%%%%%%%%

\appendix

\section{The hydro radiation code}
\label{section:code}

We have written a 1D spherical geometry, Two-Temperatures Lagrangian computer program in order to calculate the shock propagation and emitted radiation after the shock breakout. The code uses the standard von Neumann and Richtmyer staggered-mesh method to solve the equations of motion \citep{vN1960, richtmyer1967}:
\begin{subequations}
\begin{equation}
\frac{1}{\rho}\frac{\partial \rho}{\partial t} = \vec\nabla \cdot \mathbf{u},
\end{equation}
\begin{equation}
\frac{\partial u}{\partial t} = - \frac{\vec\nabla P}{\rho} - \frac{\vec\nabla U_{r}}{3/\rho}.
\end{equation}
\end{subequations}
The two equations describing the radiation and matter energy densities, are
\begin{subequations}
\begin{equation}\label{eq:de_dt}
\frac{\partial e}{\partial t} + (P+e)\vec\nabla \cdot \mathbf{u} = c\kappa_{P}\rho(U_{r}-a_{\rm{BB}}T_e^4),
\end{equation}
\begin{equation}
\frac{\partial U_{r}}{\partial t} + (\frac{4}{3}U_{r})\vec\nabla \cdot \mathbf{u} = -c\kappa_{P}\rho(U_{r}-a_{\rm{BB}}T_e^4)-\frac{\partial J}{\partial x},
\end{equation}
\end{subequations}
where, $U_r$ is the radiation energy density, $e$ is the matter energy density, $P$ is the matter pressure, $T_e$ is its temperature, $\kappa_{P}$ is Planck averaged opacity (only absorption terms are taken into account), and $J$ is the radiation flux, which is solved under the $P0$ diffusion approximation \citep{richtmyer1967, Mihalas1984, Pomraning2005}:
\begin{equation}
J = - \frac{c}{3\kappa_{R} \rho} \frac{\partial U_r}{\partial x}.
\end{equation}
Here, $\kappa_R$ is the Rosseland averaged opacity. For the equation of state (EOS) of the matter, we choose that of an ideal gas, with $\gamma=5/3$, suitable for monoatomic gas, and $\mu=0.6$ which corresponds to a fully ionized mixture of hydrogen and helium with primordial ratios. Planck opacity includes hydrogen free-free and bound-free interactions assuming primordial ratios, and Rosseland opacity includes free-free and bound-free interactions, in addition to the scattering term $\kappa_R=0.34$ cm$^2$/g which corresponds to Thomson opacity of hydrogen and helium with primordial ratios. The assumptions for the opacities and EOS are reasonable as long as hydrogen is completely ionized. Therefore, our solution is limited for temperatures higher than about $7500 ^\circ $K.

The energy equations are solved using operator splitting. First, only the hydrodynamic part is solved implicitly for both the matter and energy equations. Then, the coupling term is solved in the method described by \citet{Sapir2014}, which assures energy is conserved to numerical precision. In the last step, energy diffusion is calculated implicitly, solving a tri-diagonal equation system \citep{richtmyer1967, Mihalas1984}.

The explosion is simulated by artificially injecting thermal energy $E_{\rm exp}$ into the innermost cells as an initial condition. The flux boundary condition on the outermost cell, which is also used to determine the bolometric luminosity, is of an Eddington factor $f=0.5$ \citep{Pomraning2005} which is equivalent to Marshak boundary condition \citep{Zeldovich1967}. Since the luminosity is determined at the point $\tau \simeq c/v$, the details of the boundary condition barely affect the emission.

The code was validated via several analytic test problems, such as Elliott's extension to the Sedov-Taylor explosion which includes radiative flux \citep{Elliot1960} and Chevalier's solution for self-similar interaction of ejecta and wind \citep{Chevalier1982}, and reproduced the analytic results within numerical precision. Here, in addition, we compare our results for the planar shock breakout with previous numerical calculations.

\begin{figure}
    \centering
    \includegraphics[width=\linewidth]{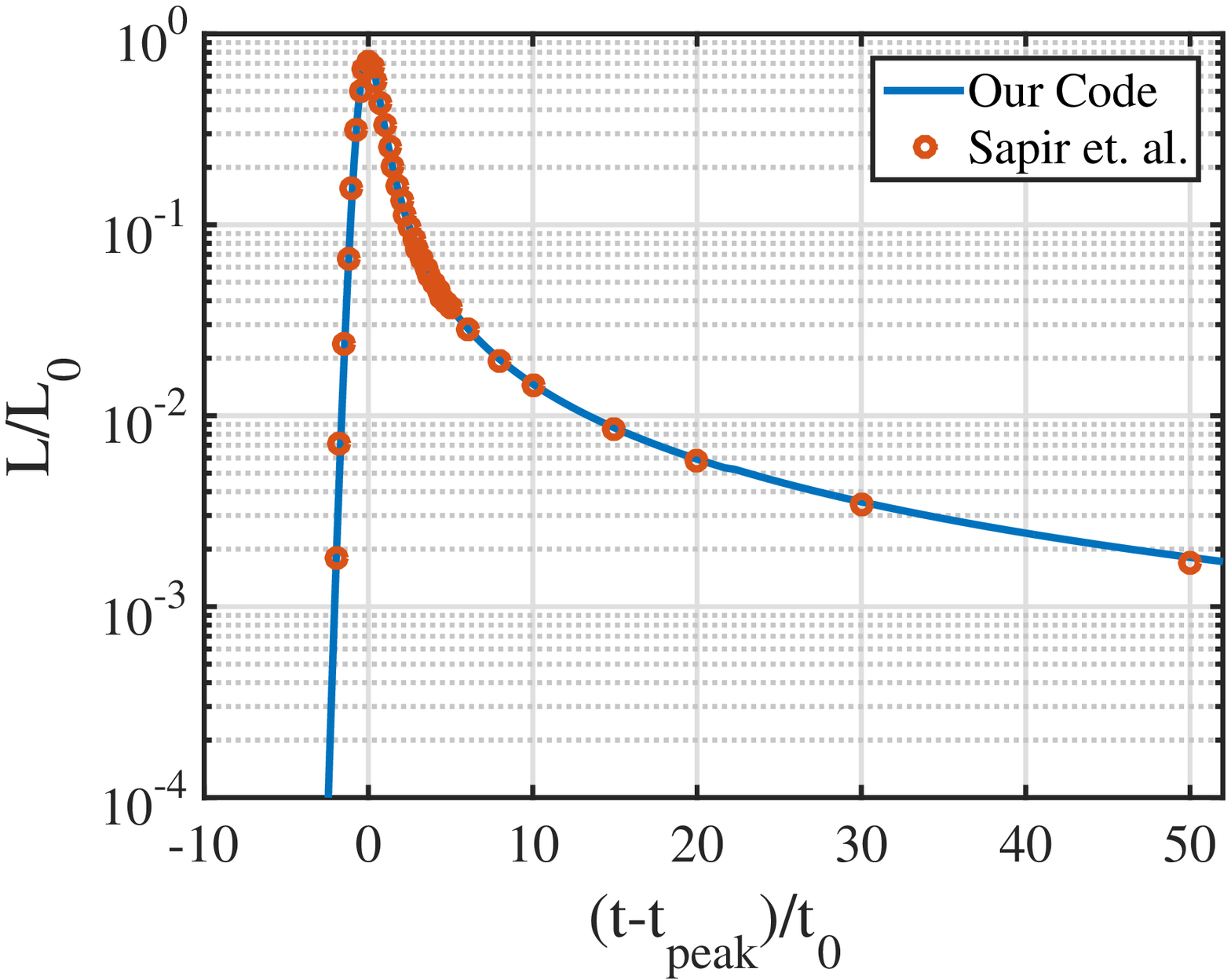}
    \caption{Normalized bolometric luminosity as a function of normalized time, for the planar shock breakout problem with $n=1.5$ \citep[see][]{Sapir2011, ginzburg2012}. The results from our code (solid blue line) fit the results from table 3 in \citet{Sapir2011} to within $1\%$ accuracy.}
    \label{fig:example_comparison}
\end{figure}

The breakout of a shock from a stellar edge in planar geometry was investigated and solved analytically by \citet{sakurai1960}. The medium is assumed to be of ideal gas with decreasing density $\rho(x) \propto x^n$ with $x$ the distance from the edge. \citet{Sapir2011} studied an extension to the problem which includes radiative flux, in the diffusion approximation. They assumed radiation dominated gas ($\gamma = 4/3$) and constant opacity, and numerically calculated a self-similar light curve emitted during the shock breakout and expansion. We use this problem as a test case for our code, and compare the light curve to \citet{Sapir2011}.

The calculation is performed in a method similar to the one described by \citet{ginzburg2012}. We use planar geometry, assume full coupling between radiation and matter, insert the appropriate density profile (with $n=1.5$) and keep only the radiation terms of the EOS. In order to simulate the explosion, we deposited thermal energy into the innermost cell as an initial condition. In figure \ref{fig:example_comparison} we present the comparison of our results with the results of \citet{Sapir2011}. An agreement to within $1\%$ is found for all times. Our light curve was normalized to the breakout values $t_0$ and $L_0$ as described in \citet{Sapir2011}.

\section{Bound-free and free-free Planck and Rosseland mean opacities before recombination}
\label{section:opacity}

\begin{figure}
    \centering
    \includegraphics[width=\linewidth]{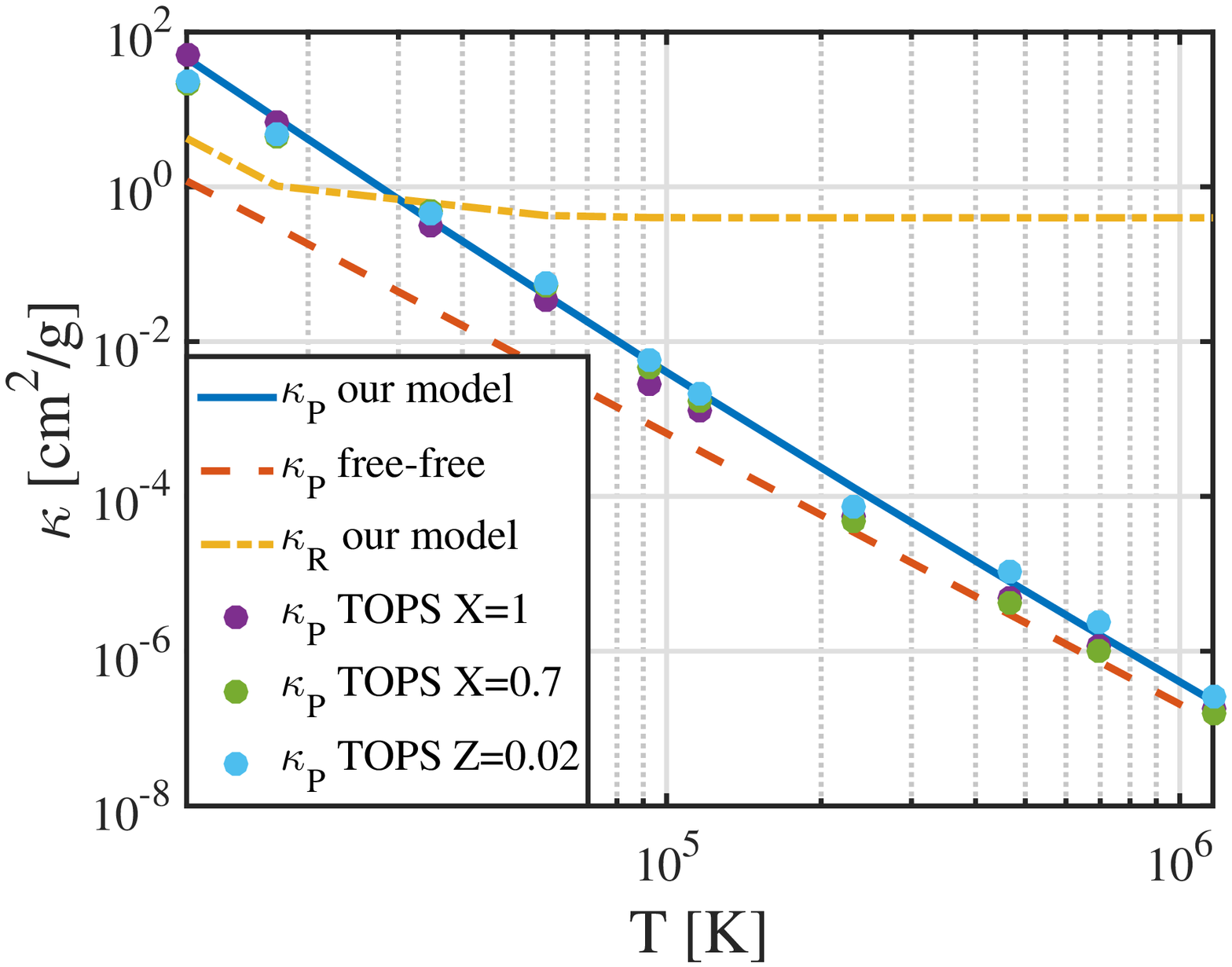}
    \caption{The opacity of different models as a function of temperature, for a typical density $\rho=10^{-10}$g/cc. The density used throughout this work for the determination of the observed temperature (solid blue line) is the Planck averaged opacity of hydrogen bound-free and free-free transitions. Planck opacity of free-free only (dashed red line) and Rosseland opacity of Thomson scattering, bound-free and free-free (dashed orange line) are plotted for comparison. In addition, bound-free and free-free Planck opacities, obtained from TOPS database, are shown for pure hydrogen (purple circles), hydrogen and helium with primordial ratios (green circles) and solar metallicity (cyan circles).}
    \label{fig:example_opacity}
\end{figure}

In the gray (P0) diffusion approximation (see appendix \ref{section:code}), radiation-matter coupling is dependent on the Planck averaged absorption opacity, while the diffusion term contains the Rosseland averaged opacity, which includes both absorption and scattering processes \citep{Pomraning2005}. In the expanding SN envelope, the absorption term is usually much smaller than the scattering term, thus the absorption opacity barely affects the bolometric luminosity (total energy flux). Nevertheless, since the absorption opacity dominates the coupling, it largely affects the observed temperature. 

Previous studies of the emission approximated the absorption opacity as hydrogen free-free dominated (e.g. NS10). In our model, we also include bound-free transitions, by approximating the cross section of each of the lowest $100$ energy levels. The exact method of approximation is specified in \citet{Zeldovich1967} chapters (V.1.3-5.).

In figure \ref{fig:example_opacity} the Planck opacity of our model is shown for various temperatures, and density $\rho=10^{-10}$g/cc (since both free-free and bound-free terms are proportional to $\rho^2$, our analysis applies for a wide range of densities [$10^{-12} - 10^{-9}$g/cc]). For comparison, the free-free term of our model, and tables from TOPS database \citep{1995ASPC...78...51M} for different mixtures are also shown. The mixtures which appear are pure hydrogen, hydrogen and helium with primordial ratios, and solar metallicity. In TOPS opacities, only bound-free and free-free transitions are considered, by artificially removing narrow lines. At breakout temperatures ($T_{\rm obs,0} \approx 1-5 \cdot 10^5 \: ^{\circ}$K) the bound-free term and the free-free term are comparable, while for lower temperatures this factor grows to more than an order of magnitude, since more photons have typical energies of the lower energy levels. For all the temperatures, the TOPS opacity changes by less than a factor of $2$ between mixtures, and is similar to our model.

We deduce that the hydrogen bound-free term determines the coupling. Since $\eta_0 \propto 1/\dot{n}$ is inversely proportional to Planck opacity, and $T_{\rm obs} \propto \eta_0^{0.07}$ (equation \ref{eq:semi_analytic_T}), inclusion of bound-free and free-free transitions from helium and metals changes the observed temperature by less than $5\%$. Neglecting hydrogen bound-free transitions, however, results in temperatures higher by $20\%$.

Figure \ref{fig:example_opacity} also depicts Rosseland opacity used in our model. For high enough temperatures, Thomson scattering indeed dominates the opacity, but at $T \approx 1 \cdot 10^4 \: ^{\circ}$K bound-free dominates. However, since Thomson scattering is density independent, and since at late times the density is much lower than $\rho=10^{-10}$g/cc, usually the scattering term is larger than the absorption term.

\section{Properties of the numerical progenitors}
\label{section:numerical_progenitors}

\input{table.tex}

% Don't change these lines
\bsp	% typesetting comment
\label{lastpage}
\end{document}

%% file: table.tex
\begin{table*}
\caption{MESA calculated progenitors (part 1)}
\label{table:numerical_progenitors}
 \begin{tabular}{llllllll}
  \hline
  $M_{\rm ZAMS}$ & $Z$ & mixing length & rotation & $M_{\rm final}$ & $M_{\rm ej}$ & $M_{\rm env}$ & $R_*$ \\
  $[M_{\odot}]$ &  & parameter & [breakup] & $[M_{\odot}]$ & $[M_{\odot}]$ & $[M_{\odot}]$ & $[R_{\odot}]$ \\
  \hline
11 & 0.02 & 2 & 0.4 & 10.07 & 9.72 & 7.19 & 518 \\ 
12 & 0.0002 & 1.5 & 0 & 12.94 & 11.39 & 8.82 & 608 \\ 
12 & 0.0002 & 1.5 & 0.2 & 12.89 & 11.27 & 8.72 & 603 \\ 
12 & 0.0002 & 2 & 0 & 12.93 & 11.31 & 8.80 & 553 \\ 
12 & 0.0002 & 2 & 0.2 & 12.88 & 11.28 & 8.68 & 552 \\ 
12 & 0.0002 & 2 & 0.4 & 12.89 & 11.19 & 8.25 & 609 \\ 
12 & 0.0002 & 3 & 0 & 12.94 & 11.31 & 8.85 & 480 \\ 
12 & 0.0002 & 3 & 0.2 & 12.93 & 11.25 & 8.54 & 512 \\ 
12 & 0.0002 & 3 & 0.4 & 12.89 & 11.25 & 8.46 & 511 \\ 
12 & 0.002 & 1.5 & 0 & 11.59 & 10.02 & 7.70 & 736 \\ 
12 & 0.002 & 1.5 & 0.2 & 10.49 & 8.86 & 5.94 & 867 \\ 
12 & 0.002 & 1.5 & 0.4 & 10.80 & 9.17 & 6.60 & 783 \\ 
12 & 0.002 & 2 & 0 & 12.23 & 10.59 & 8.20 & 614 \\ 
12 & 0.002 & 2 & 0.2 & 11.60 & 9.94 & 7.49 & 633 \\ 
12 & 0.002 & 2 & 0.4 & 11.32 & 9.64 & 6.86 & 681 \\ 
12 & 0.002 & 3 & 0 & 12.48 & 10.85 & 8.47 & 489 \\ 
12 & 0.002 & 3 & 0.2 & 11.98 & 10.41 & 7.85 & 502 \\ 
12 & 0.002 & 3 & 0.4 & 11.25 & 9.60 & 6.89 & 532 \\ 
12 & 0.002 & 5 & 0 & 12.66 & 11.02 & 8.72 & 396 \\ 
12 & 0.002 & 5 & 0.2 & 11.48 & 9.86 & 7.12 & 438 \\ 
12 & 0.002 & 5 & 0.4 & 12.23 & 10.65 & 7.90 & 432 \\ 
12 & 0.02 & 1.5 & 0 & 11.51 & 9.97 & 7.93 & 910 \\ 
12 & 0.02 & 1.5 & 0.2 & 11.36 & 9.93 & 7.74 & 926 \\ 
12 & 0.02 & 1.5 & 0.4 & 11.16 & 9.59 & 7.45 & 948 \\ 
12 & 0.02 & 1.5 & 0.6 & 9.93 & 8.26 & 5.52 & 1052 \\ 
12 & 0.02 & 2 & 0 & 11.70 & 10.07 & 8.08 & 709 \\ 
12 & 0.02 & 2 & 0.2 & 11.49 & 9.92 & 7.84 & 752 \\ 
12 & 0.02 & 2 & 0.4 & 11.23 & 9.65 & 7.48 & 778 \\ 
12 & 0.02 & 2 & 0.6 & 10.29 & 8.65 & 5.84 & 841 \\ 
12 & 0.02 & 3 & 0 & 11.92 & 10.34 & 8.36 & 559 \\ 
12 & 0.02 & 3 & 0.2 & 11.86 & 10.27 & 8.23 & 568 \\ 
12 & 0.02 & 3 & 0.4 & 11.41 & 9.81 & 7.60 & 592 \\ 
12 & 0.02 & 3 & 0.6 & 11.19 & 9.53 & 6.78 & 650 \\ 
12 & 0.02 & 5 & 0 & 12.27 & 10.70 & 8.70 & 383 \\ 
12 & 0.02 & 5 & 0.2 & 11.87 & 10.31 & 8.19 & 388 \\ 
12 & 0.02 & 5 & 0.4 & 11.78 & 10.16 & 8.11 & 390 \\ 
12 & 0.02 & 5 & 0.6 & 11.00 & 9.27 & 6.66 & 445 \\ 
12 & 0.02 & 5 & 0.8 & 10.08 & 8.27 & 4.87 & 510 \\ 
13 & 0.02 & 2 & 0 & 11.71 & 10.00 & 8.11 & 708 \\ 
13 & 0.02 & 2 & 0.2 & 11.56 & 9.95 & 7.91 & 711 \\ 
13 & 0.02 & 2 & 0.4 & 11.26 & 9.53 & 7.48 & 739 \\ 
13 & 0.02 & 2 & 0.6 & 10.01 & 8.18 & 5.50 & 847 \\ 
14 & 0.02 & 2 & 0 & 12.34 & 10.68 & 8.33 & 780 \\ 
14 & 0.02 & 2 & 0.2 & 12.04 & 10.28 & 7.89 & 817 \\ 
15 & 2e-05 & 1.5 & 0 & 14.98 & 13.27 & 10.29 & 555 \\ 
15 & 2e-05 & 1.5 & 0.4 & 14.76 & 12.74 & 9.40 & 601 \\ 
15 & 2e-05 & 3 & 0 & 14.97 & 13.07 & 10.18 & 465 \\ 
15 & 2e-05 & 3 & 0.4 & 14.91 & 12.77 & 9.59 & 496 \\ 
15 & 2e-05 & 5 & 0 & 14.98 & 13.22 & 10.30 & 390 \\ 
15 & 2e-05 & 5 & 0.4 & 14.90 & 12.85 & 9.35 & 442 \\ 
15 & 0.0002 & 1.5 & 0 & 14.95 & 13.10 & 10.13 & 608 \\ 
15 & 0.0002 & 1.5 & 0.4 & 14.77 & 12.70 & 9.40 & 658 \\ 
15 & 0.0002 & 3 & 0 & 14.92 & 13.02 & 10.02 & 476 \\ 
15 & 0.0002 & 3 & 0.4 & 14.83 & 12.79 & 9.29 & 524 \\ 
15 & 0.0002 & 5 & 0 & 14.93 & 13.16 & 10.14 & 394 \\ 
15 & 0.0002 & 5 & 0.4 & 14.79 & 12.60 & 9.27 & 448 \\ 
15 & 0.002 & 1.5 & 0 & 14.27 & 12.53 & 9.56 & 778 \\ 
15 & 0.002 & 1.5 & 0.4 & 10.37 & 8.62 & 5.18 & 894 \\ 
15 & 0.002 & 3 & 0 & 14.10 & 12.21 & 9.28 & 518 \\ 
15 & 0.002 & 3 & 0.4 & 12.62 & 10.58 & 7.41 & 562 \\ 
15 & 0.002 & 5 & 0 & 14.44 & 12.68 & 9.69 & 401 \\ 
15 & 0.002 & 5 & 0.4 & 13.63 & 11.79 & 8.45 & 432 \\ 
  \hline
 \end{tabular}
\end{table*}

\begin{table*}
\caption{MESA calculated progenitors (part 2)}
 \begin{tabular}{llllllll}
  \hline
  $M_{\rm ZAMS}$ & $Z$ & mixing length & rotation & $M_{\rm final}$ & $M_{\rm ej}$ & $M_{\rm env}$ & $R_*$ \\
  $[M_{\odot}]$ &  & parameter & [breakup] & $[M_{\odot}]$ & $[M_{\odot}]$ & $[M_{\odot}]$ & $[R_{\odot}]$ \\
  \hline
15 & 0.02 & 2 & 0 & 13.05 & 11.27 & 8.68 & 835 \\ 
15 & 0.02 & 2 & 0.2 & 12.66 & 10.94 & 8.17 & 841 \\ 
15 & 0.02 & 2 & 0.4 & 13.15 & 11.37 & 8.60 & 845 \\ 
16 & 0.02 & 2 & 0 & 14.47 & 12.68 & 9.67 & 851 \\ 
16 & 0.02 & 2 & 0.2 & 13.25 & 11.47 & 8.38 & 883 \\ 
16 & 0.02 & 2 & 0.4 & 12.71 & 10.78 & 7.66 & 943 \\ 
17 & 0.02 & 2 & 0 & 14.25 & 12.47 & 9.10 & 963 \\ 
17 & 0.02 & 2 & 0.2 & 13.44 & 11.59 & 8.07 & 977 \\ 
17 & 0.02 & 2 & 0.4 & 13.19 & 11.25 & 7.73 & 1009 \\
18 & 0.02 & 2 & 0 & 16.25 & 14.23 & 10.67 & 978 \\ 
18 & 0.02 & 2 & 0.2 & 15.22 & 13.32 & 9.61 & 1015 \\ 
18 & 0.02 & 2 & 0.4 & 13.89 & 11.93 & 8.06 & 1031 \\ 
19 & 0.02 & 2 & 0 & 15.46 & 13.53 & 9.58 & 1046 \\ 
19 & 0.02 & 2 & 0.2 & 14.47 & 12.50 & 8.34 & 1077 \\ 
20 & 2e-05 & 3 & 0.4 & 17.88 & 17.17 & 10.91 & 666 \\ 
20 & 0.0002 & 1.5 & 0 & 19.94 & 17.41 & 13.33 & 471 \\ 
20 & 0.0002 & 1.5 & 0.4 & 19.48 & 16.86 & 11.38 & 812 \\ 
20 & 0.0002 & 3 & 0 & 19.91 & 17.90 & 12.93 & 616 \\ 
20 & 0.0002 & 3 & 0.4 & 19.45 & 16.76 & 11.12 & 680 \\ 
20 & 0.0002 & 5 & 0 & 19.92 & 17.57 & 13.11 & 511 \\ 
20 & 0.0002 & 5 & 0.4 & 19.38 & 16.76 & 11.02 & 561 \\ 
20 & 0.002 & 1.5 & 0 & 14.69 & 12.62 & 7.83 & 1027 \\ 
20 & 0.002 & 1.5 & 0.4 & 12.73 & 10.59 & 5.18 & 991 \\ 
20 & 0.002 & 3 & 0 & 15.70 & 13.67 & 8.65 & 703 \\ 
20 & 0.002 & 3 & 0.4 & 14.72 & 12.44 & 7.04 & 735 \\ 
20 & 0.002 & 5 & 0 & 16.70 & 14.69 & 9.94 & 529 \\ 
20 & 0.02 & 2 & 0 & 15.41 & 13.38 & 9.11 & 1025 \\ 
20 & 0.02 & 2 & 0.2 & 15.00 & 13.02 & 8.63 & 1019 \\ 
21 & 0.02 & 2 & 0 & 15.74 & 13.70 & 9.04 & 1037 \\ 
21 & 0.02 & 2 & 0.2 & 14.98 & 12.95 & 8.26 & 1030 \\ 
21 & 0.02 & 2 & 0.4 & 11.47 & 9.18 & 4.43 & 925 \\ 
22 & 0.02 & 2 & 0 & 17.37 & 15.01 & 10.36 & 1039 \\ 
22 & 0.02 & 2 & 0.4 & 12.31 & 10.23 & 4.88 & 945 \\ 
23 & 0.02 & 2 & 0 & 16.80 & 14.70 & 9.40 & 1032 \\ 
23 & 0.02 & 2 & 0.2 & 13.09 & 11.04 & 5.60 & 986 \\ 
23 & 0.02 & 2 & 0.4 & 12.94 & 10.80 & 5.13 & 942 \\ 
24 & 0.02 & 2 & 0 & 16.32 & 14.22 & 8.47 & 1040 \\ 
24 & 0.02 & 2 & 0.2 & 14.63 & 12.55 & 6.75 & 1005 \\ 
25 & 2e-05 & 1.5 & 0 & 24.97 & 22.17 & 16.01 & 154 \\ 
25 & 2e-05 & 1.5 & 0.4 & 19.00 & 15.87 & 7.78 & 1018 \\ 
25 & 0.0002 & 3 & 0.4 & 18.33 & 15.18 & 7.53 & 784 \\ 
25 & 0.0002 & 5 & 0 & 24.90 & 22.58 & 15.95 & 149 \\ 
25 & 0.0002 & 5 & 0.4 & 16.20 & 13.22 & 5.38 & 639 \\ 
25 & 0.002 & 1.5 & 0 & 14.30 & 11.63 & 5.21 & 994 \\ 
25 & 0.002 & 1.5 & 0.4 & 20.00 & 17.31 & 10.06 & 1030 \\ 
25 & 0.002 & 3 & 0 & 17.94 & 15.31 & 8.78 & 799 \\ 
25 & 0.002 & 3 & 0.4 & 19.40 & 16.51 & 9.39 & 756 \\ 
25 & 0.002 & 5 & 0 & 18.44 & 16.22 & 9.29 & 646 \\ 
25 & 0.002 & 5 & 0.4 & 20.54 & 17.79 & 10.83 & 640 \\ 
25 & 0.02 & 2 & 0 & 16.08 & 13.89 & 7.86 & 1016 \\ 
25 & 0.02 & 2 & 0.2 & 13.62 & 11.46 & 5.34 & 950 \\ 
26 & 0.02 & 2 & 0 & 16.45 & 14.10 & 7.79 & 943 \\ 
26 & 0.02 & 2 & 0.2 & 13.88 & 11.49 & 5.22 & 900 \\ 
27 & 0.02 & 2 & 0 & 16.22 & 13.38 & 7.14 & 917 \\ 
27 & 0.02 & 2 & 0.2 & 13.37 & 10.61 & 4.28 & 811 \\ 
28 & 0.02 & 2 & 0 & 15.66 & 12.81 & 6.12 & 844 \\ 
28 & 0.02 & 2 & 0.2 & 19.53 & 16.63 & 10.24 & 902 \\ 
29 & 0.02 & 2 & 0 & 16.21 & 13.40 & 6.25 & 790 \\ 
30 & 0.02 & 2 & 0 & 16.05 & 13.37 & 5.62 & 703 \\ 
30 & 0.02 & 2 & 0.2 & 15.34 & 12.52 & 5.04 & 697 \\ 
35 & 0.02 & 2 & 0 & 17.11 & 14.47 & 4.55 & 377 \\ 
35 & 0.02 & 2 & 0.2 & 17.10 & 14.37 & 4.55 & 380 \\
  \hline
 \end{tabular}
\end{table*}

%% file: main.bbl
\begin{thebibliography}{}
\makeatletter
\relax
\def\mn@urlcharsother{\let\do\@makeother \do\$\do\&\do\#\do\^\do\_\do\%\do\~}
\def\mn@doi{\begingroup\mn@urlcharsother \@ifnextchar [ {\mn@doi@}
  {\mn@doi@[]}}
\def\mn@doi@[#1]#2{\def\@tempa{#1}\ifx\@tempa\@empty \href
  {http://dx.doi.org/#2} {doi:#2}\else \href {http://dx.doi.org/#2} {#1}\fi
  \endgroup}
\def\mn@eprint#1#2{\mn@eprint@#1:#2::\@nil}
\def\mn@eprint@arXiv#1{\href {http://arxiv.org/abs/#1} {{\tt arXiv:#1}}}
\def\mn@eprint@dblp#1{\href {http://dblp.uni-trier.de/rec/bibtex/#1.xml}
  {dblp:#1}}
\def\mn@eprint@#1:#2:#3:#4\@nil{\def\@tempa {#1}\def\@tempb {#2}\def\@tempc
  {#3}\ifx \@tempc \@empty \let \@tempc \@tempb \let \@tempb \@tempa \fi \ifx
  \@tempb \@empty \def\@tempb {arXiv}\fi \@ifundefined
  {mn@eprint@\@tempb}{\@tempb:\@tempc}{\expandafter \expandafter \csname
  mn@eprint@\@tempb\endcsname \expandafter{\@tempc}}}

\bibitem[\protect\citeauthoryear{{Anderson} et~al.,}{{Anderson}
  et~al.}{2014}]{Anderson2014}
{Anderson} J.~P.,  et~al., 2014, \mn@doi [\apj] {10.1088/0004-637X/786/1/67},
  \href {http://adsabs.harvard.edu/abs/2014ApJ...786...67A} {786, 67}

\bibitem[\protect\citeauthoryear{{Arcavi} et~al.,}{{Arcavi}
  et~al.}{2012}]{Arcavi2012}
{Arcavi} I.,  et~al., 2012, \mn@doi [\apjl] {10.1088/2041-8205/756/2/L30},
  \href {http://adsabs.harvard.edu/abs/2012ApJ...756L..30A} {756, L30}

\bibitem[\protect\citeauthoryear{{Blinnikov}, {Eastman}, {Bartunov},
  {Popolitov}  \& {Woosley}}{{Blinnikov} et~al.}{1998}]{Blinnikov1998}
{Blinnikov} S.~I.,  {Eastman} R.,  {Bartunov} O.~S.,  {Popolitov} V.~A.,
  {Woosley} S.~E.,  1998, \mn@doi [\apj] {10.1086/305375}, \href
  {http://adsabs.harvard.edu/abs/1998ApJ...496..454B} {496, 454}

\bibitem[\protect\citeauthoryear{{Chevalier}}{{Chevalier}}{1976}]{Chevalier1976}
{Chevalier} R.~A.,  1976, \mn@doi [\apj] {10.1086/154557}, \href
  {http://adsabs.harvard.edu/abs/1976ApJ...207..872C} {207, 872}

\bibitem[\protect\citeauthoryear{{Chevalier}}{{Chevalier}}{1982}]{Chevalier1982}
{Chevalier} R.~A.,  1982, \mn@doi [\apj] {10.1086/160126}, \href
  {http://adsabs.harvard.edu/abs/1982ApJ...258..790C} {258, 790}

\bibitem[\protect\citeauthoryear{{Chevalier}}{{Chevalier}}{1992}]{Chevalier1992}
{Chevalier} R.~A.,  1992, \mn@doi [\apj] {10.1086/171612}, \href
  {http://adsabs.harvard.edu/abs/1992ApJ...394..599C} {394, 599}

\bibitem[\protect\citeauthoryear{{Colgate}}{{Colgate}}{1974}]{Colgate1974}
{Colgate} S.~A.,  1974, \mn@doi [\apj] {10.1086/152632}, \href
  {http://adsabs.harvard.edu/abs/1974ApJ...187..333C} {187, 333}

\bibitem[\protect\citeauthoryear{{Dessart}, {Hillier}, {Waldman}  \&
  {Livne}}{{Dessart} et~al.}{2013}]{Dessart2013}
{Dessart} L.,  {Hillier} D.~J.,  {Waldman} R.,   {Livne} E.,  2013, \mn@doi
  [\mnras] {10.1093/mnras/stt861}, \href
  {http://adsabs.harvard.edu/abs/2013MNRAS.433.1745D} {433, 1745}

\bibitem[\protect\citeauthoryear{Elliott}{Elliott}{1960}]{Elliot1960}
Elliott L.~A.,  1960, \mn@doi [Proceedings of the Royal Society of London A:
  Mathematical, Physical and Engineering Sciences] {10.1098/rspa.1960.0188},
  258, 287

\bibitem[\protect\citeauthoryear{{Ensman} \& {Burrows}}{{Ensman} \&
  {Burrows}}{1992}]{Ensman1992}
{Ensman} L.,  {Burrows} A.,  1992, \mn@doi [\apj] {10.1086/171542}, \href
  {http://adsabs.harvard.edu/abs/1992ApJ...393..742E} {393, 742}

\bibitem[\protect\citeauthoryear{{Falk}}{{Falk}}{1978}]{Falk1978}
{Falk} S.~W.,  1978, \mn@doi [\apjl] {10.1086/182810}, \href
  {http://adsabs.harvard.edu/abs/1978ApJ...225L.133F} {225, L133}

\bibitem[\protect\citeauthoryear{{Faran} et~al.,}{{Faran}
  et~al.}{2014}]{Faran2014}
{Faran} T.,  et~al., 2014, \mn@doi [\mnras] {10.1093/mnras/stu955}, \href
  {http://adsabs.harvard.edu/abs/2014MNRAS.442..844F} {442, 844}

\bibitem[\protect\citeauthoryear{{Gall} et~al.,}{{Gall}
  et~al.}{2015}]{Gall2015}
{Gall} E.~E.~E.,  et~al., 2015, \mn@doi [\aap] {10.1051/0004-6361/201525868},
  \href {http://adsabs.harvard.edu/abs/2015A%26A...582A...3G} {582, A3}

\bibitem[\protect\citeauthoryear{{Gezari} et~al.,}{{Gezari}
  et~al.}{2008}]{Gezari2008}
{Gezari} S.,  et~al., 2008, \mn@doi [\apjl] {10.1086/591647}, \href
  {http://adsabs.harvard.edu/abs/2008ApJ...683L.131G} {683, L131}

\bibitem[\protect\citeauthoryear{{Ginzburg} \& {Balberg}}{{Ginzburg} \&
  {Balberg}}{2012}]{ginzburg2012}
{Ginzburg} S.,  {Balberg} S.,  2012, \mn@doi [\apj]
  {10.1088/0004-637X/757/2/178}, \href
  {http://adsabs.harvard.edu/abs/2012ApJ...757..178G} {757, 178}

\bibitem[\protect\citeauthoryear{{Glebbeek}, {Gaburov}, {de Mink}, {Pols}  \&
  {Portegies Zwart}}{{Glebbeek} et~al.}{2009}]{2009A&A...497..255G}
{Glebbeek} E.,  {Gaburov} E.,  {de Mink} S.~E.,  {Pols} O.~R.,   {Portegies
  Zwart} S.~F.,  2009, \mn@doi [\aap] {10.1051/0004-6361/200810425}, \href
  {http://adsabs.harvard.edu/abs/2009A%26A...497..255G} {497, 255}

\bibitem[\protect\citeauthoryear{{Gonz{\'a}lez-Gait{\'a}n}
  et~al.,}{{Gonz{\'a}lez-Gait{\'a}n} et~al.}{2015}]{Gonzales2015}
{Gonz{\'a}lez-Gait{\'a}n} S.,  et~al., 2015, \mn@doi [\mnras]
  {10.1093/mnras/stv1097}, \href
  {http://adsabs.harvard.edu/abs/2015MNRAS.451.2212G} {451, 2212}

\bibitem[\protect\citeauthoryear{{Grassberg}, {Imshennik}  \&
  {Nadyozhin}}{{Grassberg} et~al.}{1971}]{Grassberg1971}
{Grassberg} E.~K.,  {Imshennik} V.~S.,   {Nadyozhin} D.~K.,  1971, \mn@doi
  [\apss] {10.1007/BF00654604}, \href
  {http://adsabs.harvard.edu/abs/1971Ap%26SS..10...28G} {10, 28}

\bibitem[\protect\citeauthoryear{{Imshennik}, {Nadezhin}  \&
  {Utrobin}}{{Imshennik} et~al.}{1981}]{Imshennik1981}
{Imshennik} V.~S.,  {Nadezhin} D.~K.,   {Utrobin} V.~P.,  1981, \mn@doi [\apss]
  {10.1007/BF00654026}, \href
  {http://adsabs.harvard.edu/abs/1981Ap%26SS..78..105I} {78, 105}

\bibitem[\protect\citeauthoryear{{Katz}, {Budnik}  \& {Waxman}}{{Katz}
  et~al.}{2010}]{Katz2010}
{Katz} B.,  {Budnik} R.,   {Waxman} E.,  2010, \mn@doi [\apj]
  {10.1088/0004-637X/716/1/781}, \href
  {http://adsabs.harvard.edu/abs/2010ApJ...716..781K} {716, 781}

\bibitem[\protect\citeauthoryear{{Katz}, {Sapir}  \& {Waxman}}{{Katz}
  et~al.}{2012}]{Katz2012}
{Katz} B.,  {Sapir} N.,   {Waxman} E.,  2012, \mn@doi [\apj]
  {10.1088/0004-637X/747/2/147}, \href
  {http://adsabs.harvard.edu/abs/2012ApJ...747..147K} {747, 147}

\bibitem[\protect\citeauthoryear{{Magee} et~al.,}{{Magee}
  et~al.}{1995}]{1995ASPC...78...51M}
{Magee} N.~H.,  et~al., 1995, in {Adelman} S.~J.,  {Wiese} W.~L.,  eds,
  Astronomical Society of the Pacific Conference Series Vol. 78, Astrophysical
  Applications of Powerful New Databases. p.~51

\bibitem[\protect\citeauthoryear{{Matzner} \& {McKee}}{{Matzner} \&
  {McKee}}{1999}]{Matzner1999}
{Matzner} C.~D.,  {McKee} C.~F.,  1999, \mn@doi [\apj] {10.1086/306571}, \href
  {http://adsabs.harvard.edu/abs/1999ApJ...510..379M} {510, 379}

\bibitem[\protect\citeauthoryear{Mihalas \& Mihalas}{Mihalas \&
  Mihalas}{1984}]{Mihalas1984}
Mihalas D.,  Mihalas B.~W.,  1984, Foundations of Radiation Hydrodynamics.
ed. B. W. Mihalas and D. Mihalas

\bibitem[\protect\citeauthoryear{{Morozova}, {Piro}, {Renzo}  \&
  {Ott}}{{Morozova} et~al.}{2016}]{Morozova2016}
{Morozova} V.,  {Piro} A.~L.,  {Renzo} M.,   {Ott} C.~D.,  2016, preprint,
  \href {http://adsabs.harvard.edu/abs/2016arXiv160308530M} {} (\mn@eprint
  {arXiv} {1603.08530})

\bibitem[\protect\citeauthoryear{{Nakar} \& {Sari}}{{Nakar} \&
  {Sari}}{2010}]{Nakar2010}
{Nakar} E.,  {Sari} R.,  2010, \mn@doi [\apj] {10.1088/0004-637X/725/1/904},
  \href {http://adsabs.harvard.edu/abs/2010ApJ...725..904N} {725, 904}

\bibitem[\protect\citeauthoryear{{Nieuwenhuijzen} \& {de
  Jager}}{{Nieuwenhuijzen} \& {de Jager}}{1990}]{1990A&A...231..134N}
{Nieuwenhuijzen} H.,  {de Jager} C.,  1990, \aap, \href
  {http://adsabs.harvard.edu/abs/1990A%26A...231..134N} {231, 134}

\bibitem[\protect\citeauthoryear{{Nugis} \& {Lamers}}{{Nugis} \&
  {Lamers}}{2000}]{2000A&A...360..227N}
{Nugis} T.,  {Lamers} H.~J.~G.~L.~M.,  2000, \aap, \href
  {http://adsabs.harvard.edu/abs/2000A%26A...360..227N} {360, 227}

\bibitem[\protect\citeauthoryear{{Paxton}, {Bildsten}, {Dotter}, {Herwig},
  {Lesaffre}  \& {Timmes}}{{Paxton} et~al.}{2011}]{2011ApJS..192....3P}
{Paxton} B.,  {Bildsten} L.,  {Dotter} A.,  {Herwig} F.,  {Lesaffre} P.,
  {Timmes} F.,  2011, \mn@doi [\apjs] {10.1088/0067-0049/192/1/3}, \href
  {http://adsabs.harvard.edu/abs/2011ApJS..192....3P} {192, 3}

\bibitem[\protect\citeauthoryear{{Paxton} et~al.,}{{Paxton}
  et~al.}{2013}]{2013ApJS..208....4P}
{Paxton} B.,  et~al., 2013, \mn@doi [\apjs] {10.1088/0067-0049/208/1/4}, \href
  {http://adsabs.harvard.edu/abs/2013ApJS..208....4P} {208, 4}

\bibitem[\protect\citeauthoryear{{Paxton} et~al.,}{{Paxton}
  et~al.}{2015}]{2015ApJS..220...15P}
{Paxton} B.,  et~al., 2015, \mn@doi [\apjs] {10.1088/0067-0049/220/1/15}, \href
  {http://adsabs.harvard.edu/abs/2015ApJS..220...15P} {220, 15}

\bibitem[\protect\citeauthoryear{{Piro}, {Chang}  \& {Weinberg}}{{Piro}
  et~al.}{2010}]{Piro2010}
{Piro} A.~L.,  {Chang} P.,   {Weinberg} N.~N.,  2010, \mn@doi [\apj]
  {10.1088/0004-637X/708/1/598}, \href
  {http://adsabs.harvard.edu/abs/2010ApJ...708..598P} {708, 598}

\bibitem[\protect\citeauthoryear{Pomraning}{Pomraning}{2005}]{Pomraning2005}
Pomraning G.~C.,  2005, The Equations of Radiation Hydrodynamics.
Dover Publications, INC

\bibitem[\protect\citeauthoryear{{Rabinak} \& {Waxman}}{{Rabinak} \&
  {Waxman}}{2011}]{Rabinak2011}
{Rabinak} I.,  {Waxman} E.,  2011, \mn@doi [\apj] {10.1088/0004-637X/728/1/63},
  \href {http://adsabs.harvard.edu/abs/2011ApJ...728...63R} {728, 63}

\bibitem[\protect\citeauthoryear{Richtmyer \& Morton}{Richtmyer \&
  Morton}{1967}]{richtmyer1967}
Richtmyer R.~D.,  Morton K.~W.,  1967, Difference Methods for Initial-Value
  Problems, 2 edn.
Interscience Publishers

\bibitem[\protect\citeauthoryear{{Rubin} et~al.,}{{Rubin}
  et~al.}{2015}]{Rubin2015}
{Rubin} A.,  et~al., 2015, preprint, \href
  {http://adsabs.harvard.edu/abs/2015arXiv151200733R} {} (\mn@eprint {arXiv}
  {1512.00733})

\bibitem[\protect\citeauthoryear{{Sakurai}}{{Sakurai}}{1960}]{sakurai1960}
{Sakurai} A.,  1960, Communications on Pure and Applied Mathematics, 13, 353

\bibitem[\protect\citeauthoryear{{Sanders} et~al.,}{{Sanders}
  et~al.}{2015}]{Sanders2015}
{Sanders} N.~E.,  et~al., 2015, \mn@doi [\apj] {10.1088/0004-637X/799/2/208},
  \href {http://adsabs.harvard.edu/abs/2015ApJ...799..208S} {799, 208}

\bibitem[\protect\citeauthoryear{{Sapir} \& {Halbertal}}{{Sapir} \&
  {Halbertal}}{2014}]{Sapir2014}
{Sapir} N.,  {Halbertal} D.,  2014, \mn@doi [\apj]
  {10.1088/0004-637X/796/2/145}, \href
  {http://adsabs.harvard.edu/abs/2014ApJ...796..145S} {796, 145}

\bibitem[\protect\citeauthoryear{{Sapir} \& {Waxman}}{{Sapir} \&
  {Waxman}}{2016}]{Sapir2016}
{Sapir} N.,  {Waxman} E.,  2016, preprint, \href
  {http://adsabs.harvard.edu/abs/2016arXiv160703700S} {} (\mn@eprint {arXiv}
  {1607.03700})

\bibitem[\protect\citeauthoryear{{Sapir}, {Katz}  \& {Waxman}}{{Sapir}
  et~al.}{2011}]{Sapir2011}
{Sapir} N.,  {Katz} B.,   {Waxman} E.,  2011, \mn@doi [\apj]
  {10.1088/0004-637X/742/1/36}, \href
  {http://adsabs.harvard.edu/abs/2011ApJ...742...36S} {742, 36}

\bibitem[\protect\citeauthoryear{{Schawinski} et~al.,}{{Schawinski}
  et~al.}{2008}]{Schawinski2008}
{Schawinski} K.,  et~al., 2008, \mn@doi [Science] {10.1126/science.1160456},
  \href {http://adsabs.harvard.edu/abs/2008Sci...321..223S} {321, 223}

\bibitem[\protect\citeauthoryear{Sedov}{Sedov}{1959}]{Sedov1959}
Sedov L.~I.,  1959, Similarity and Dimensional Methods in Machanics.
ed. L. I. Sedov

\bibitem[\protect\citeauthoryear{{Shigeyama}, {Nomoto}  \&
  {Hashimoto}}{{Shigeyama} et~al.}{1988}]{Shigeyama1988}
{Shigeyama} T.,  {Nomoto} K.,   {Hashimoto} M.,  1988, \aap, \href
  {http://adsabs.harvard.edu/abs/1988A%26A...196..141S} {196, 141}

\bibitem[\protect\citeauthoryear{{Smartt}}{{Smartt}}{2009}]{Smartt2009}
{Smartt} S.~J.,  2009, \mn@doi [\araa] {10.1146/annurev-astro-082708-101737},
  \href {http://adsabs.harvard.edu/abs/2009ARA%26A..47...63S} {47, 63}

\bibitem[\protect\citeauthoryear{{Taylor}}{{Taylor}}{1950}]{Taylor1950}
{Taylor} G.,  1950, Royal Society of London Proceedings Serias A, 251

\bibitem[\protect\citeauthoryear{{Tominaga}, {Blinnikov}, {Baklanov},
  {Morokuma}, {Nomoto}  \& {Suzuki}}{{Tominaga} et~al.}{2009}]{Tominaga2009}
{Tominaga} N.,  {Blinnikov} S.,  {Baklanov} P.,  {Morokuma} T.,  {Nomoto} K.,
  {Suzuki} T.,  2009, \mn@doi [\apjl] {10.1088/0004-637X/705/1/L10}, \href
  {http://adsabs.harvard.edu/abs/2009ApJ...705L..10T} {705, L10}

\bibitem[\protect\citeauthoryear{{Tominaga}, {Morokuma}, {Blinnikov},
  {Baklanov}, {Sorokina}  \& {Nomoto}}{{Tominaga} et~al.}{2011}]{Tominaga2011}
{Tominaga} N.,  {Morokuma} T.,  {Blinnikov} S.~I.,  {Baklanov} P.,  {Sorokina}
  E.~I.,   {Nomoto} K.,  2011, \mn@doi [\apjs] {10.1088/0067-0049/193/1/20},
  \href {http://adsabs.harvard.edu/abs/2011ApJS..193...20T} {193, 20}

\bibitem[\protect\citeauthoryear{{Vink}, {de Koter}  \& {Lamers}}{{Vink}
  et~al.}{2001}]{2001A&A...369..574V}
{Vink} J.~S.,  {de Koter} A.,   {Lamers} H.~J.~G.~L.~M.,  2001, \mn@doi [\aap]
  {10.1051/0004-6361:20010127}, \href
  {http://adsabs.harvard.edu/abs/2001A%26A...369..574V} {369, 574}

\bibitem[\protect\citeauthoryear{{Weaver}}{{Weaver}}{1976}]{Weaver1976}
{Weaver} T.~A.,  1976, \mn@doi [\apjs] {10.1086/190398}, \href
  {http://adsabs.harvard.edu/abs/1976ApJS...32..233W} {32, 233}

\bibitem[\protect\citeauthoryear{{Woosley}}{{Woosley}}{1988}]{Woosley1988}
{Woosley} S.~E.,  1988, \mn@doi [\apj] {10.1086/166468}, \href
  {http://adsabs.harvard.edu/abs/1988ApJ...330..218W} {330, 218}

\bibitem[\protect\citeauthoryear{Zel'dovich \& Raizer}{Zel'dovich \&
  Raizer}{1967}]{Zeldovich1967}
Zel'dovich Y.~B.,  Raizer Y.~P.,  1967, Physics of Shock Waves and
  High-Temperature Hydrodynamic Phenomena, 2 edn.
Academic Press Inc.

\bibitem[\protect\citeauthoryear{{von Neumann} \& {Richmyer}}{{von Neumann} \&
  {Richmyer}}{1960}]{vN1960}
{von Neumann} J.,  {Richmyer} R.~D.,  1960, Communications on Pure and Applied
  Mathematics, 13, 353

\makeatother
\end{thebibliography}
